\documentclass[aps,floats]{revtex4-2}
\usepackage{amsmath,amssymb}
\usepackage{graphicx,epsfig}
\usepackage[greek,english]{babel}
\usepackage{bbold}

\begin{document}
\bibliographystyle {plain}

\pdfoutput=1
\def\oppropto{\mathop{\propto}} 
\def\opsimeq{\mathop{\simeq}}
\def\opoverderline{\mathop{\overline}}
\def\operarrow{\mathop{\longrightarrow}}
\def\opsim{\mathop{\sim}}

\def\opmin{\mathop{\min}} 
\def\opmax{\mathop{\max}} 
\def\oplim{\mathop{\lim}}

\title{  Convergence properties of Markov models for image generation \\
with applications to spin-flip dynamics and to diffusion processes} 


\author{C\'ecile Monthus}
\affiliation{Universit\'e Paris-Saclay, CNRS, CEA, Institut de Physique Th\'eorique, 91191 Gif-sur-Yvette, France}


\begin{abstract}
In the field of Markov models for image generation, the main idea is to learn how non-trivial images are gradually destroyed by a trivial forward Markov dynamics over the large time window $[0,t]$ converging towards pure noise for $t \to + \infty$, and to implement the non-trivial backward time-dependent Markov dynamics over the same time window $[0,t]$ starting from pure noise at $t$ in order to generate new images at time $0$. The goal of the present paper is to analyze the convergence properties of this reconstructive backward dynamics as a function of the time $t$ using the spectral properties of the trivial continuous-time forward dynamics for the $N$ pixels $n=1,..,N$. The general framework is applied to two cases : (i) when each pixel $n$ has only two states $S_n=\pm 1$ with Markov jumps between them; (ii) when each pixel $n$ is characterized by a continuous variable $x_n$ that diffuses on an interval $]x_-,x_+[$ that can be either finite or infinite.

\end{abstract}

\maketitle


\section{ Introduction }

Among the myriad of algorithms that have appeared recently in the area of image generation,
the generative Markov models (see the review \cite{GeneReview} and references therein)
are extremely interesting from the point of view of non-equilibrium statistical physics,
as stressed from various perspectives in the recent works \cite{gene_noneq,Biroli_largedim,Ambrogioni1,Ambrogioni2,Ambrogioni3,Biroli_dynamical,emergence,pathintegral,GeneNonEq,wasserstein,garrahan,Ambrogioni4,Ambrogioni5,c_generativeLargedevPixels,cumulants}.
Indeed, the forward dynamics where non-trivial images are gradually destroyed towards pure noise
is related to the issues concerning the transient non-equilibrium relaxation
towards equilibrium when starting from arbitrary initial conditions, 
while the backward reconstructive dynamics 
is based on the idea of reversing the time-arrow in Markov processes.
Of course this backward Markov dynamics is governed by a very artificial time-dependent generator
that would never occur in real life but that can be implemented on a computer : 
a simple metaphor of the method would be that you film how a sandcastle built on the beach
 is gradually washed away by the tide,
and then you look at the film backward in time to see how the sandcastle is progressively reconstructed.
Besides all the essential issues concerning their optimal implementations for real data,
the generative Markov models also raise a lot of interesting theoretical questions
in order to better understand their properties
in relation with other areas of statistical physics and of stochastic processes
\cite{gene_noneq,Biroli_largedim,Ambrogioni1,Ambrogioni2,Ambrogioni3,Biroli_dynamical,emergence,pathintegral,GeneNonEq,wasserstein,garrahan,Ambrogioni4,Ambrogioni5,c_generativeLargedevPixels,cumulants}.

The goal of the present paper is to address the following limited theoretical question :
if the empirical knowledge from the many interesting non-trivial images has been gathered 
into an initial probability distribution $P^{ini}(\vec C)$ for the global configuration 
$\vec C=(C_1,..,C_N)$ of the $N$ pixels $n=1,2,..,N$, 
where the configuration $C_n$ of the pixel $n$ is either discrete or continuous,
what are the convergence properties of the forward dynamics during the time-window $[0,t]$,
i.e. what are the time-scales associated to the relaxations towards zero of the non-trivial pixel-correlations 
that are present in the initial condition $P^{ini}(\vec C) $ ?
And then what are the convergence properties of the reconstructive backward dynamics 
towards the initial distribution $P^{ini}(\vec C)$ as a function of the time $t$?

The paper is organized as follows.
The general spectral properties of the forward dynamics are discussed in section \ref{sec_forward}
and are then used to study the reconstructive backward dynamics in section \ref{sec_backward}.
Finally, this general framework is applied to specific models, 
namely to spin models with spin-flip dynamics in section \ref{sec_spins}
and to diffusion processes in section \ref{sec_diff}.
Our conclusions are summarized in section \ref{sec_conclusion}.


\section{ Convergence properties of the forward Markov dynamics }

\label{sec_forward}

In this section, the general properties of the forward Markov dynamics
for the global configuration $\vec C=(C_1,..,C_N)$ of the $N$ pixels $n=1,2,..,N$
are described using the very convenient bra-ket notations familiar from quantum physics.

\subsection{ Factorized form of the global forward propagator over the $N$ pixels}

The forward dynamics should converge towards some simple steady state representing 'noise'
that can be easily generated.
 The simplest choice is when the $N$ pixels evolve towards 'noise' independently,
 i.e. when the continuous-time Markov generator $W$ for the whole configuration  $\vec C=(C_1,..,C_N)$
is simply the sum of identical elementary generators $w_n$ acting on the individual pixels
  \begin{eqnarray}
W=\sum_{n=1}^N w_n
\label{Wsum}
\end{eqnarray}
Then the propagator $\langle \vec c \vert e^{  t W} \vert \vec C \rangle $ between 
the global configuration $\vec C=(C_1,..,C_N)$ at the initial time $t=0$ and 
the global configuration $\vec c=(c_1,..,c_N)$ at the time $t$ 
reduces to the product of the $N$ propagators $\langle c_n \vert e^{  t w_n} \vert C_n \rangle$ 
associated to the $N$ pixels $n=1,..,N$
  \begin{eqnarray}
\langle \vec c \vert e^{  t W} \vert \vec C \rangle
=   \langle c_1,..,c_N \vert \ \ e^{ \displaystyle t \sum_{n=1}^N w_n} \ \ \vert C_1,..,C_N \rangle
= \prod_{n=1}^N  \langle c_n \vert e^{  t w_n} \vert C_n \rangle
\label{factorizedpropagator}
\end{eqnarray}
 So one just needs to analyse the propagator $\langle c \vert e^{  t w} \vert C \rangle $ associated to a single pixel, as described in the next section.


\subsection{ Spectral properties of the forward propagator $\langle c \vert e^{  t w} \vert C \rangle $ associated to a single pixel}

To simplify the notations in the present general analysis, we will note $(1+\alpha_{max} )$
the number of possible configurations $c$ for a single pixel, 
while the applications to spin models corresponding to the simplest case $\alpha_{max}= 1$ 
and to diffusion models with only discrete spectra corresponding to $\alpha_{max}=+\infty$
will be considered in the sections \ref{sec_spins} and \ref{sec_diff} respectively.
Here the goal is to highlight the general spectral properties independently of the technical details 
that are specific to Markov jump processes on discrete spaces or to diffusion processes on continuous spaces.

\subsubsection{ Spectral decomposition in the bi-orthogonal basis of left and right eigenvectors of the generator $w$ }

The spectral decomposition of the propagator $\langle c \vert e^{  t w} \vert C \rangle $ associated to a single pixel
  \begin{eqnarray}
\langle c \vert e^{  t w} \vert C \rangle 
&& = \sum_{\alpha=0}^{\alpha_{max}} e^{-t e_{\alpha} } \langle c \vert r_{\alpha} \rangle \langle  l_{\alpha} \vert C \rangle
 \nonumber \\ && 
= p_*(c)+\sum_{\alpha=1}^{\alpha_{max}} e^{-t e_{\alpha} }r_{\alpha} (c) l_{\alpha}(C)
\label{spectralpixel}
\end{eqnarray}
involves the $\alpha_{max} $
non-vanishing eigenvalues $(-e_{\alpha} ) \ne 0 $ of the Markov generator $w$ labelled by $\alpha=1,2,..,\alpha_{max}$
with their corresponding right and left eigenvectors 
  \begin{eqnarray}
 w \vert r_{\alpha} \rangle && =  - e_{\alpha}   \vert r_{\alpha} \rangle 
\nonumber \\
\langle  l_{\alpha} \vert w  && =  - e_{\alpha}   \langle  l_{\alpha} \vert
\label{eigenw}
\end{eqnarray}
while for $\alpha=0$, the vanishing eigenvalue $e_0=0$ is associated
 to the trivial left eigenvector $l_0 (C)  =  1 $ corresponding to the conservation of probability
and to the right eigenvector $r_0 (c)  =  p_*(C)$ corresponding to the steady state
  \begin{eqnarray}
   l_0 (C) && =  1
\nonumber \\
    r_0 (c) && =  p_*(C) 
\label{eigenw0}
\end{eqnarray}
with the bi-orthonormalization
  \begin{eqnarray}
  \delta_{\alpha,\beta} = \langle  l_{\alpha} \vert r_{\beta} \rangle = \sum_c  \langle  l_{\alpha} \vert c \rangle
  \langle c \vert r_{\beta} \rangle
\label{biortho}
\end{eqnarray}
The spectral decomposition of Eq. \ref{spectralpixel} describes the relaxation towards the steady state $p_*(c)$ for any initial condition $C$: the right eigenvectors $r_{\alpha \ne 0} (c) $ are the relaxation modes,
while the left eigenvectors $l_{\alpha \ne 0} (c) $ are the observables 
that decay towards zero with a single exponential $e^{-t e_{\alpha} } $
  \begin{eqnarray}
 l_{\alpha}^{av}(t \vert  C,0) \equiv 
   \sum_c \langle  l_{\alpha} \vert c \rangle \langle  c \vert e^{  t w} \vert  C \rangle
 = \langle  l_{\alpha} \vert e^{  t w} \vert  C \rangle  
 = e^{-t e_{\alpha} }    l_{\alpha}(C ) 
\label{leftobservable}
\end{eqnarray}
while the dynamics of other observables involve several or the whole spectrum of eigenvalues $ e_{\alpha \ne 0}$,
since any observable can be expanded on the basis of the left eigenvectors $l_{\alpha}$.
For a general Markov generator $w$, 
the eigenvalues $e_{\alpha \ne 0} $ can be complex with positive real parts and can produce oscillations,
while they are always real in the presence of detailed-balance, as recalled in the next subsection.


\subsubsection{ Taking into account the additional property of detailed-balance }

As explained in textbooks \cite{gardiner,vankampen,risken} 
and in specific applications to various Markov models (see for instance \cite{glauber,Felderhof,siggia,kimball,peschel,jpb_antoine,pierre,texier,us_eigenvaluemethod,Castelnovo,c_pearson,c_boundarydriven,c_DBsusy}), the generators of Markov processes satisfying detailed-balance
are related to quantum Hamiltonians via similarity transformations.

In Markov generative models, the chosen forward dynamics 
usually satisfies this additional property of detailed-balance,
where the steady current $J_*(c,c')$ produced by the steady state $p_*(.)$
vanishes between any pair $c \ne c'$ of configurations 
 \begin{eqnarray}
0 = J_*(c,c') \equiv \langle c \vert w \vert c' \rangle p_*(c') -  \langle c' \vert w \vert c \rangle p_*(c)
\label{DBdef}
\end{eqnarray}
It is then useful to perform the following similarity transformation from the Markov generator $w$
towards the real symmetric quantum Hamiltonian $H$ with the matrix elements
\begin{eqnarray}
\langle c \vert H \vert C \rangle
\equiv  - \frac{1}{  \sqrt{p_* (c) }}  \langle c \vert w \vert C \rangle    \sqrt{p_* (C) } 
=  - \frac{1}{  \sqrt{p_* (C) } }  \langle C \vert w \vert c \rangle  \sqrt{p_* (c) } = \langle C \vert H \vert c \rangle
\label{similarity}
\end{eqnarray}
The propagator $\langle c \vert e^{ - t H} \vert C \rangle $ associated to the quantum Hamiltonian 
then reads in terms of the Markov propagator of Eq. \ref{spectralpixel}
  \begin{eqnarray}
 \langle c \vert e^{ - t H} \vert C \rangle  =
   \frac{1}{  \sqrt{p_* (c) }}  \langle c \vert e^{  t w} \vert C \rangle    \sqrt{p_* (C) } 
&&  =   \sum_{\alpha=0}^{\alpha_{max}} e^{-t e_{\alpha} } 
  \left[\frac{r_{\alpha}(c) }{  \sqrt{p_* (c) }} \right] \left[   l_{\alpha} (C) \sqrt{p_* (C) } \right]
  \nonumber \\
&& \equiv \sum_{\alpha=0}^{\alpha_{max}} e^{-t e_{\alpha} }  \psi_{\alpha} (c)  \psi_{\alpha} (C)
\label{spectralH}
\end{eqnarray}
with the following conclusions :

(i) the eigenvalues $(-e_{\alpha})$ of the Markov generator $w$ 
involve the real eigenvalues $e_{\alpha}$ of the Hamiltonian $H$
  \begin{eqnarray}
    e_0  =0< e_1< e_2<...<e_{\alpha_{max}}
\label{ealphaordered}
\end{eqnarray}

(ii)  the corresponding real eigenstates $\psi_{\alpha}$ of the real symmetric Hamiltonian $H$
  \begin{eqnarray}
 H \vert \psi_{\alpha} \rangle  =   e_{\alpha}   \vert \psi_{\alpha} \rangle 
\label{eigenH}
\end{eqnarray}
satisfying the orthonormalization
  \begin{eqnarray}
  \delta_{\alpha,\beta} = \langle  \psi_{\alpha} \vert \psi_{\beta} \rangle = \sum_c  \langle  \psi_{\alpha} \vert c \rangle
  \langle c \vert \psi_{\beta} \rangle
\label{ortho}
\end{eqnarray}
are related to the left and to the right eigenvectors of the Markov generator via
\begin{eqnarray}
 \psi_{\alpha} (c) = \frac{r_{\alpha}(c) }{  \sqrt{p_* (c) }} =  l_{\alpha} (c) \sqrt{p_* (c) }
\label{eigenwpsi}
\end{eqnarray}

It is then often more convenient to work only with the left eigenvectors $l_{\alpha}$ :
the translation of
the bi-orthonormalization of Eq. \ref{biortho} or of the quantum orthonormalization of Eq. \ref{ortho}
  \begin{eqnarray}
  \delta_{\alpha,\beta}   = \sum_c   l_{\alpha} (c)  l_{\beta}(c) p_*(c)
\label{ortholeft}
\end{eqnarray}
means that the left eigenvectors $l_{\alpha} $ form an orthonormal family 
with respect to the steady state measure $p_*(c)$,
while the spectral decomposition of the propagator of Eq. \ref{spectralpixel}
becomes
  \begin{eqnarray}
\langle c \vert e^{  t w} \vert C \rangle 
&& =p_*(c) \sum_{\alpha=0}^{\alpha_{max}} e^{-t e_{\alpha} } l_{\alpha} (c)   l_{\alpha} (C)
 \nonumber \\ && 
= p_*(c) \left[ 1+\sum_{\alpha=1}^{\alpha_{max}} e^{-t e_{\alpha} }l_{\alpha} (c) l_{\alpha}(C) \right]
\label{spectralpixelleft}
\end{eqnarray}
 

\subsection{ Properties of of the global forward propagator for the $N$ pixels }

\subsubsection{ Spectral decomposition of the global forward propagator $\langle \vec c \vert e^{  t W} \vert \vec C \rangle $  }

Plugging the spectral decomposition of Eq. \ref{spectralpixelleft}
for the individual propagators $\langle c_n \vert e^{  t w_n} \vert C_n \rangle $ associated to the $N$ pixels $n=1,..,N$ 
into the global propagator of Eq. \ref{factorizedpropagator}
yields
  \begin{eqnarray}
\langle \vec c \vert e^{  t W} \vert \vec C \rangle
= \prod_{n=1}^N  \langle c_n \vert e^{  t w_n} \vert C_n \rangle
&& = \prod_{n=1}^N \left[ p_*(c_n) \sum_{\alpha_n=0}^{\alpha_{max}} e^{-t e_{\alpha_n} }l_{\alpha_n} (c_n) l_{\alpha_n}(C_n)\right]
\nonumber \\
&& = P_*( \vec c) \left[ 1+ \sum_{\vec \alpha \ne \vec 0} e^{- t E_{\vec \alpha} } L_{\vec \alpha}(\vec c) L_{\vec \alpha}(\vec C) \right]
\label{factorizedpropagatorspectral}
\end{eqnarray}
that characterizes the convergence towards the global factorized steady state
  \begin{eqnarray}
P_*( \vec c) \equiv  \prod_{n=1}^N p_*(c_n) 
\label{Psteadyglobal}
\end{eqnarray}
The global energies $E_{\vec \alpha=(\alpha_1,..,\alpha_N)} $ reduce to the sum of individual energies $e_{\alpha_n} $
  \begin{eqnarray}
E_{\vec \alpha} = \sum_{n=1}^N e_{\alpha_n}
\label{Esum}
\end{eqnarray}
while the global left eigenvectors reduce to product of individual left eigenvectors 
  \begin{eqnarray}
  L_{\vec \alpha} (\vec C)  = \prod_{n=1}^N l_{\alpha_n}(C_n) 
  \label{Lprod}
\end{eqnarray}
that form an orthonormal family 
with respect to the steady state measure $P_*(\vec c)$ 
as a consequence of the same property of Eq. \ref{ortholeft}
concerning a single pixel
  \begin{eqnarray}
  \delta_{\vec \alpha,\vec \beta} 
  = \sum_{\vec c}   L_{\vec \alpha} (\vec c)  L_{\vec \beta}(\vec c) P_*( \vec c)
\label{ortholeftN}
\end{eqnarray}


\subsubsection{ Leading asymptotic exponential decay governed by the first excited energy $e_1$ }

The sum formula of Eq. \ref{Esum} mean that there are a lot of degeneracies.
In particular, the first strictly positive energy $e_1$ is obtained when only a single coefficient is unity $\alpha_{i}=1$
while the $(N-1)$ other vanish $\alpha_{n\ne i}=0 $ 
  \begin{eqnarray}
e_1=E_{ (\alpha_n=\delta_{n,i})} 
\label{Efirstexcited}
\end{eqnarray}
It is thus $N$-times degenerate with the choice of the excited component $i \in \{1,..,N\}$,
while the other $(N-1)$ components $n \ne i$ are in the ground state $e_0=0$,
so that the corresponding left eigenvectors  
  \begin{eqnarray}
  L_{(\alpha_n=\delta_{n,i})} (\vec C)  = l_1(C_i) \prod_{n\ne i} l_{0}(C_n) =  l_1(C_i)
  \label{LfirstExcited}
\end{eqnarray}
reduce to the one-pixel observables $l_1(C_i)$.

As a consequence, the leading contribution of order $e^{-t e_1}$ governing
the convergence of Eq. \ref{factorizedpropagatorspectral} towards the steady state
  \begin{eqnarray}
\langle \vec c \vert e^{  t W} \vert \vec C \rangle
 = P_*( \vec c) \bigg[ 1+  e^{- t e_1 } \sum_{i=1}^N l_1(c_i) l_1(C_i) +...\bigg]
\label{factorizedpropagatorspectralonlyleftfirstexcited}
\end{eqnarray}
only involve these $N$ observables $l_1(.)$ associated to the $N$ individual pixels,
while observables involving more pixels will appear in higher-order terms,
as shown by the dynamics of Eq. \ref{leftobservableGlobal} for the observables associated 
to the left eigenvectors $L_{\vec \alpha}$ that form a basis of observables.


\subsubsection{ Global convergence as measured by the Kemeny time }

In order to characterize the global convergence towards the steady state $P_*(\vec C)$ of the $N$ pixels,
another interesting characteristic time is the Kemeny time (see \cite{c_kemeny} and references therein for more details)
  \begin{eqnarray}
\tau^{Kemeny}_N = \tau^{Spectral}_N = \tau^{Space}_N 
\label{kemeny}
\end{eqnarray}
with its spectral definition $\tau^{Spectral}_N $
and its real-space definition $\tau^{Space}_N  $
that can be written for our present framework as follows :

(i) The spectral Kemeny time $\tau^{Spectral}_N  $ 
is defined as the sum of the inverses of the non-vanishing eigenvalues $E_{\vec \alpha  \ne \vec 0} $ 
 \begin{eqnarray}
\tau^{Spectral}_N 
\equiv  \sum_{\vec \alpha \ne \vec 0} \frac{1}{E_{\vec \alpha}} = \int_0^{+\infty} dt  Z_N(t) 
\label{tauspectral}
\end{eqnarray}
and thus can be rewritten as the integral over the time $t$ of the spectral partition function
 \begin{eqnarray}
Z_N(t) \equiv   \sum_{\vec \alpha \ne \vec 0} e^{- t E_{\vec \alpha}} 
= \sum_{\vec \alpha } e^{- t E_{\vec \alpha}} -1
\label{SpectralPartitionZ}
\end{eqnarray}
Since the global energy $E_{\vec \alpha} $ of Eq. \ref{Esum}
 reduces to the sum of individuals energies $e_{\alpha_n} $, 
 the global spectral 
partition function of Eq. \ref{SpectralPartitionZ} reads
 \begin{eqnarray}
Z_N(t) \equiv   \sum_{\vec \alpha } \left[ \prod_{n=1}^N e^{- t e_{\alpha_n}} \right] -1 
=   \left[ \sum_{\alpha} e^{- t e_{\alpha}} \right]^N -1
=    \left[ 1+\sum_{\alpha \ne 0} e^{- t e_{\alpha}} \right]^N -1 
=  \left[ 1+Z_1(t) \right]^N -1
\label{SpectralPartitionZz}
\end{eqnarray}
in terms of the spectral partition function $Z_1(t)$ associated to a single pixel
 \begin{eqnarray}
Z_1(t) \equiv  \sum_{\alpha \ne 0} e^{- t e_{\alpha}} = \sum_{\alpha=1}^{\alpha_{max}} e^{- t e_{\alpha}}
\label{SpectralPartitionz1}
\end{eqnarray}
Plugging Eq. \ref{SpectralPartitionZz} into the Kemeny time of Eq. \ref{tauspectral}
yields the dependence with respect to the number $N$ of pixels
 \begin{eqnarray}
\tau^{Spectral}_N =   \int_0^{+\infty} dt  Z_N(t) = \int_0^{+\infty} dt  \left(  \left[ 1+Z_1(t) \right]^N -1\right) 
\label{tauspectralN}
\end{eqnarray}

When the number $(1+\alpha_{max})$ of configurations for a single pixel is finite,
the asymptotic behavior for large $N$ of the integral of Eq. \ref{tauspectralN}
can be evaluated via the saddle-point method at the lower boundary $t=0$ where $Z_1(t) $ is maximum.
 The series expansion at first order in $t$
 \begin{eqnarray}
Z_1(t) = \sum_{\alpha=1}^{\alpha_{max}} \left[ 1- t e_{\alpha} +O(t^2) \right] = \alpha_{max} -t 
\sum_{\alpha=1}^{\alpha_{max}}  e_{\alpha} +O(t^2) \equiv Z_1(0)-t Z_1'(0)  +O(t^2)
\label{SpectralPartitionz1series}
\end{eqnarray}
 leads to the following exponential growth in $N$ 
as the size $\left[ 1+ \alpha_{max}\right]^N$ of the configuration space of the $N$ pixels
 \begin{eqnarray}
\tau^{Spectral}_N \opsimeq_{N \to + \infty}   \frac{\left[ 1+Z_1(0) \right]^{N+1}}{N \left[ -Z_1'(0) \right]]}
=  \frac{\left[ 1+ \alpha_{max}\right]^{N+1}}{ \displaystyle N \left[ \sum_{\alpha=1}^{\alpha_{max}}  e_{\alpha}\right] }
\label{tauspectralNlarge}
\end{eqnarray}

(ii) Remarkably, when the Mean-First-Passage-Time $\tau(\vec c \vert \vec C)$ at configuration $\vec c$ 
when starting at configuration $\vec C$
is averaged over the final configuration $\vec c$ drawn with the steady state $ P_*(\vec c)$,
the result is actually independent of the initial configuration $\vec C$
and represents the real-space definition $\tau_N^{Space}$ of the Kemeny time 
(see \cite{c_kemeny} and references therein for more details)
 \begin{eqnarray}
  \sum_{\vec c} P_{*}(\vec c) \tau(\vec c, \vec C) \equiv \tau^{Space}_N \ \ \ \ { \rm independent \ \ of } \ \ \vec C
\label{taukemeny}
\end{eqnarray}

The correspondences between the spectral and the real-space perspectives \cite{c_kemeny,c_SkewDB}
is also useful to write the Mean-First-Passage-Time $\tau(\vec c \vert \vec C)$ itself
in terms of the spectral properties
\cite{c_SkewDB}
 \begin{eqnarray}
   \tau(\vec c \vert \vec C) = \frac{1 }{P_*(\vec c)} \sum_{\vec \alpha \ne \vec 0} R_{\vec \alpha} (\vec c)  \frac{1}{E_{\vec \alpha}}
   \left[ L_{\vec \alpha} (\vec c) - L_{\vec \alpha} (\vec C)\right]
\label{MFTPspectral}
\end{eqnarray}
that can be simplified 
in the presence of the detailed-balance property $R_{\vec \alpha} (\vec c)=P_*(\vec c) L_{\vec \alpha} (\vec c)$
into the following expression
that involves only the excited eigenvalues $E_{\vec \alpha \ne \vec 0} $ 
and the excited left eigenvectors $L_{\vec \alpha \ne \vec 0} $
 \begin{eqnarray}
   \tau(\vec c \vert \vec C) =  \sum_{\vec \alpha \ne \vec 0}  \frac{ L_{\vec \alpha} (\vec c)  \left[ L_{\vec \alpha} (\vec c) - L_{\vec \alpha} (\vec C)\right] }{E_{\vec \alpha}}
\label{MFTPspectralleft}
\end{eqnarray}



\subsection{ Application of the trivial forward propagator $\langle \vec c \vert e^{  t W} \vert \vec C \rangle$ to the non-trivial initial condition $P^{ini}( \vec C)$ }

The application of the forward propagator $\langle \vec c \vert e^{  t W} \vert \vec C \rangle $ of Eq. \ref{factorizedpropagatorspectral} to the non-trivial initial condition $P^{ini}( \vec C)$
produces the probability $P( \vec c, t) $ to see the global configuration $\vec c$ at time $t$ 
  \begin{eqnarray}
P( \vec c, t) && =  \sum_{ \vec C}  \langle \vec c \vert e^{  t W} \vert \vec C \rangle P^{ini}(\vec C) 
= \langle \vec c \vert e^{  t W} \vert  P^{ini} \rangle
\nonumber \\
&& =   P_*( \vec c) \left[ 1+ \sum_{\vec \alpha \ne \vec 0} e^{- t E_{\vec \alpha} } L_{\vec \alpha}(\vec c)  L^{ini}_{\vec \alpha} \right]
\label{ptforward}
\end{eqnarray}
where $L^{ini}_{\vec \alpha} $ represent the observables associated
 to the left eigenvector $L_{\vec \alpha} $ of Eq. \ref{Lprod}
in the initial distribution $P^{ini}(\vec C) $ 
 \begin{eqnarray}
 L^{ini}_{\vec \alpha} \equiv  \langle L_{\vec \alpha} \vert P^{ini} \rangle 
 = \sum_{ \vec C} L_{\vec \alpha}(\vec C)  P^{ini}(\vec C)
 = \sum_{ \vec C} \left[ \prod_{n=1}^N l_{\alpha_n}(C_n)\right]  P^{ini}(\vec C)
\label{Leftini}
\end{eqnarray}
that can be used to rewrite the initial distribution $P^{ini}(\vec C) $ corresponding to the special case $t=0$ in Eq. \ref{ptforward}
  \begin{eqnarray}
P^{ini}(\vec C)&& =   P_*( \vec C) \left[ 1+ \sum_{\vec \alpha \ne \vec 0}  L_{\vec \alpha}(\vec C)  L^{ini}_{\vec \alpha} \right]
\label{LeftiniPiniexpansion}
\end{eqnarray}

As a consequence of Eq. \ref{ortholeftN},
the relaxation towards zero of the observables $ L_{\vec \alpha}^{av}(t )   $ associated
 to the left eigenvectors $L_{\vec \alpha \ne \vec 0} $
 involve a single global energy $E_{\vec \alpha} $ of Eq. \ref{Esum}
  \begin{eqnarray}
 L_{\vec \alpha}^{av}(t )   \equiv \sum_{\vec c} L_{\alpha}(\vec c) P( \vec c, t)
 =  L^{ini}_{\alpha} e^{- t E_{\vec \alpha} }
 = L^{ini}_{\alpha} \ \ e^{ \displaystyle  - t \sum_{n=1}^N e_{\alpha_n} }
\label{leftobservableGlobal}
\end{eqnarray}

In summary, the non-trivial information contained in the initial distribution $P^{ini}(\vec C) $
has been decomposed into the basis of observables $L_{\vec \alpha \ne \vec 0}^{av}(t ) $ 
that start at their initial values $ L^{ini}_{\vec \alpha} $ of Eq. \ref{Leftini}
and that relax towards zero with the corresponding global energies $E_{\vec \alpha} $.


\section{ Convergence properties of the reconstructive backward dynamics }

\label{sec_backward}

In this section, the convergences properties of the reconstructive backward dynamics
are studied using the spectral properties of the forward dynamics described in the previous section.

\subsection{ Backward propagator $B(  \vec C, 0 \vert \vec c, t) $
 in terms of the forward propagator 
 and the initial condition $P^{ini}(\vec C)$ }

Within the forward dynamics described in the previous section,
the probability $P( \vec c, t) $ to see the global configuration $\vec c$ at time $t$
 was already discussed in Eq. \ref{ptforward}, while
the joint probability $P^{Joint}(\vec c_2, t_2 ; \vec c_1, t_1 ) $ 
to see the configuration $\vec c_1$ at time $t_1$ 
and the configuration $\vec c_2$ at time $t_2$
with $0\leq t_1<t_2 \leq t$ reads in terms of the generator $W$ and the initial condition $P^{ini} $
  \begin{eqnarray}
P^{Joint}(\vec c_2, t_2 ; \vec c_1, t_1 ) = \langle \vec c_2 \vert e^{  (t_2-t_1) W} \vert \vec c_1 \rangle P( \vec c_1, t_1)
=  \langle \vec c_2 \vert e^{  (t_2-t_1) W} \vert \vec c_1 \rangle 
\langle \vec c_1 \vert e^{  t_1 W} \vert \vec P^{ini} \rangle
\label{jointforward12}
\end{eqnarray}

The backward propagator $B(  \vec c_1, t_1 \vert \vec c_2, t_2) $ can be defined 
as the  
conditional probability that appears when the joint probability of Eq. \ref{jointforward12} is analyzed backwards in time 
  \begin{eqnarray}
P^{Joint}(\vec c_2, t_2 ; \vec c_1, t_1 ) = B(  \vec c_1, t_1 \vert \vec c_2, t_2) P( \vec c_2, t_2)
= B(  \vec c_1, t_1 \vert \vec c_2, t_2)
\langle \vec c_2 \vert e^{  t_2 W} \vert \vec P^{ini} \rangle
\label{jointbackward12}
\end{eqnarray}
that leads to the expression
  \begin{eqnarray}
B(  \vec c_1, t_1 \vert \vec c_2, t_2)  = \frac{ P^{Joint}(\vec c_2, t_2 ; \vec c_1, t_1 ) }{ P( \vec c_2, t_2) }
&& = \frac{ \langle \vec c_2 \vert e^{  (t_2-t_1) W} \vert \vec c_1 \rangle P( \vec c_1, t_1) }{ P( \vec c_2, t_2) }
\nonumber \\
&& =  \frac{ \langle \vec c_2 \vert e^{  (t_2-t_1) W} \vert \vec c_1 \rangle \langle \vec c_1 \vert e^{  t_1 W} \vert \vec P^{ini} \rangle }{ \langle \vec c_2 \vert e^{  t_2 W} \vert  P^{ini} \rangle }
\label{backwardpropagator12}
\end{eqnarray}
The effective time-dependent backward generator that depends on $P^{ini} $ can be then obtained by considering the infinitesimal time difference $t_2-t_1=dt$, either for Markov jump processes or for diffusion processes,
 but will not be discussed in the present paper where
we will focus on the backward propagator $B(  \vec C, 0 \vert \vec c, t )$
corresponding to the case $t_1=0$ and $t_2=t$
  \begin{eqnarray}
 B(  \vec C, 0 \vert \vec c, t ) = \frac{ P^{Joint}(\vec c, t ; \vec C, 0 ) }{  P( \vec c, t) }
  =  \frac{ \langle \vec c \ \vert e^{  t W} \vert \vec C \rangle P^{ini}(\vec C) }
 { \langle \vec c \ \vert e^{  t W} \vert  P^{ini} \rangle  }
 =  \frac{ \langle \vec c \ \vert e^{  t W} \vert \vec C \rangle P^{ini}(\vec C) }
 { \displaystyle \sum_{ \vec C'}  \langle \vec c \ \vert e^{  t W} \vert \vec C' \rangle P^{ini}(\vec C')  }
\label{backwardpropagator}
\end{eqnarray}
in order to analyze its properties as a function of the time $t$.

In the limit of infinite time $t=+\infty$, the joint distribution $P^{Joint}(\vec c, t ; \vec C, 0 ) $ reduces to the product
of the steady state $P_*( \vec c) $ and of the initial distribution $P^{ini}(\vec C) $
  \begin{eqnarray}
P^{Joint}(\vec c, t=+\infty ; \vec C, 0 ) = P_*( \vec c) P^{ini}(\vec C)
\label{jointforwardinfty}
\end{eqnarray}
so that the backward propagator of Eq. \ref{backwardpropagator}
  \begin{eqnarray}
 B(  \vec C, 0 \vert \vec c, t=\infty ) = \frac{ P_*( \vec c) P^{ini}(\vec C) }{ P_*( \vec c)  }
 =  P^{ini}(\vec C)
\label{backwardpropagatorinfty}
\end{eqnarray}
reduces to the initial distribution $P^{ini}(\vec C)$ for any $\vec c$.

To analyze what happens for finite time $t$,
it is useful to plug the spectral decomposition of the forward propagator 
$\langle \vec c \ \vert e^{  t W} \vert \vec C \rangle$
of Eq. \ref{factorizedpropagatorspectral} into Eq. \ref{backwardpropagator}
to obtain
  \begin{eqnarray}
 B(  \vec C, 0 \vert \vec c, t ) 
&& =  \frac{ P_*( \vec c) \bigg[ 1+ \sum_{\vec \alpha \ne \vec 0} e^{- t E_{\vec \alpha} } L_{\vec \alpha}(\vec c) L_{\vec \alpha}(\vec C) \bigg] P^{ini}(\vec C) }
 { \displaystyle \sum_{ \vec C'}  P_*( \vec c) \bigg[ 1+ \sum_{\vec \beta \ne \vec 0} e^{- t E_{\vec \beta} } L_{\vec \beta}(\vec c) L_{\vec \beta}(\vec C') \bigg] P^{ini}(\vec C')  }
 \nonumber \\
&& =  \frac{ \displaystyle \left[ 1 + \sum_{\vec \alpha \ne \vec 0} e^{- t E_{\vec \alpha} } 
L_{\vec \alpha}(\vec c)  L_{\vec \alpha}(\vec C) \right] P^{ini}(\vec C)}
 { \displaystyle 1 + \sum_{\vec \beta \ne \vec 0} e^{- t E_{\vec \beta} } L_{\vec \beta}(\vec c) 
 L^{ini}_{\vec \beta}  } 
\label{backwardpropagatorspectral}
\end{eqnarray}
where $L^{ini}_{\vec \beta} = \langle L_{\vec \beta} \vert  P^{ini} \rangle$ have been discussed in Eq. \ref{Leftini}.


\subsection{ Application of the finite-time backward propagator $B(  \vec C, 0 \vert \vec c, t ) $ to the steady state $P_*( \vec c)$ }

The application of the finite-time backward-propagator $ B(  \vec C, 0 \vert \vec c, t )  $ of Eqs \ref{backwardpropagator}
and \ref{backwardpropagatorspectral}
to the steady state $P_*( \vec c)$ 
leads to the reconstructed distribution
  \begin{eqnarray}
B^{[0,t]}(\vec C) && \equiv \sum_{\vec c}  B(  \vec C, 0 \vert \vec c, t )   P_*(\vec c)
 = \sum_{\vec c} P_*(\vec c) \frac{ \langle \vec c \ \vert e^{  t W} \vert \vec C \rangle P^{ini}(\vec C) }
 { \langle \vec c \ \vert e^{  t W} \vert P^{ini} \rangle  }
\nonumber \\
&& = P^{ini}(\vec C) \sum_{\vec c}   P_*(\vec c) \ 
 \frac{ \displaystyle   1 + \sum_{\vec \alpha \ne \vec 0} e^{- t E_{\vec \alpha} } 
L_{\vec \alpha}(\vec c)  L_{\vec \alpha}(\vec C)  }
 { \displaystyle 1 + \sum_{\vec \beta \ne \vec 0} e^{- t E_{\vec \beta} } L_{\vec \beta}(\vec c) 
 L^{ini}_{\vec \beta}  } 
\label{B0reconstructedbackward}
\end{eqnarray}
that converges towards the initial distribution $P^{ini}(\vec C) $
in the infinite-time limit 
  \begin{eqnarray}
B^{[0,t=\infty]}(\vec C) 
=  P^{ini}(\vec C) \sum_{\vec c}   P_*(\vec c) =  P^{ini}(\vec C)
\label{B0reconstructedbackwardinfty}
\end{eqnarray}

To analyze this convergence,
 it is convenient to write the difference between the two probability distributions
  \begin{eqnarray}
B^{[0,t]}(\vec C) -  P^{ini}(\vec C) 
&& = P^{ini}(\vec C) \sum_{\vec c}   P_*(\vec c) \left[ 
 \frac{ \displaystyle   1 + \sum_{\vec \alpha \ne \vec 0} e^{- t E_{\vec \alpha} } 
L_{\vec \alpha}(\vec c)  L_{\vec \alpha}(\vec C)  }
 { \displaystyle 1 + \sum_{\vec \beta \ne \vec 0} e^{- t E_{\vec \beta} } L_{\vec \beta}(\vec c) 
 L^{ini}_{\vec \beta}  } -1 \right]
 \nonumber \\
 && =P^{ini}(\vec C)  \sum_{\vec c}   P_*(\vec c)
\left(  \frac{ \displaystyle     \sum_{\vec \alpha \ne \vec 0} e^{- t E_{\vec \alpha} } 
L_{\vec \alpha}(\vec c)  \left[ L_{\vec \alpha}(\vec C) -   L^{ini}_{\vec \alpha}  \right] }
 { \displaystyle 1 + \sum_{\vec \beta \ne \vec 0} e^{- t E_{\vec \beta} } L_{\vec \beta}(\vec c) 
 L^{ini}_{\vec \beta}  }  \right)
\label{B0reconstructedbackwarddifference}
\end{eqnarray}


\subsection{ Leading asymptotic exponential convergence of $ B^{[0,t]}(\vec C)$ towards $P^{ini}(\vec C) $ }

If one keeps only the contributions associated to the first excited energy $e_1$ of Eq. \ref{Efirstexcited}
with the corresponding $N$ left eigenvectors of Eq.  \ref{LfirstExcited}
both in the numerator and denominator of Eq. \ref{B0reconstructedbackwarddifference},
one obtains
   \begin{eqnarray}
&& B^{[0,t]}(\vec C) -  P^{ini}(\vec C) 
  \opsimeq_{t \to + \infty} P^{ini}(\vec C)   \sum_{\vec c}   P_*(\vec c)
\left(  \frac{ \displaystyle e^{- t e_1 } 
\sum_{i=1}^N  l_1(c_i) \left[  l_1(C_i) -  \langle l_1^{[i]} \vert  P^{ini} \rangle  \right] +...}
 { \displaystyle 1 + \displaystyle e^{- t e_1 } 
\sum_{j=1}^N  l_1(c_j) \langle l_1^{[j]} \vert  P^{ini} \rangle +...  }  \right) 
\nonumber \\
&&  \opsimeq_{t \to + \infty} P^{ini}(\vec C)    \sum_{\vec c} \left[\prod_{n=1}^N p_*(c_n) \right]
e^{- t e_1 } 
\sum_{i=1}^N  l_1(c_i) \left[  l_1(C_i) -  \langle l_1^{[i]} \vert  P^{ini} \rangle \right]
\left(   1 - \displaystyle e^{- t e_1 } 
\sum_{j=1}^N  l_1(c_j)  \langle l_1^{[j]} \vert  P^{ini} \rangle    \right) +...
\label{B0reconstructedbackwarddifferencecalcul}
\end{eqnarray}
where the observables 
that remain from the global observables $L^{ini}_{\vec \beta} = \langle L_{\vec \beta} \vert  P^{ini} \rangle$
of Eq. \ref{Leftini} are the single-pixel-observables $l_1^{[i]} $ associated to the $N$ pixels $i=1,..,N$.
 \begin{eqnarray}
 \langle l_1^{[i]} \vert  P^{ini} \rangle = \sum_{ \vec C} l_1(C_i)  P^{ini}(\vec C)
\label{Leftinil1i}
\end{eqnarray}

The orthogonality between the steady state $p_*(c_i)=r_0(c_i)$ and 
the first excited left eigenvector $l_1(c_i)$ of Eq. \ref{biortho} yields
that the contribution of order $e^{- t e_1 } $ in Eq. \ref{B0reconstructedbackwarddifferencecalcul}
vanishes, 
while in the contribution of order $e^{- t 2 e_1 } $,
only the terms $i=j$ survive and involve $\sum_{c_i}  p_*(c_i) l_1^2(c_i) = \langle l_1 \vert r_1 \rangle =1$,
so that the final leading contribution
of Eq. \ref{B0reconstructedbackwarddifferencecalcul}
is given by
  \begin{eqnarray}
 B^{[0,t]}(\vec C) -  P^{ini}(\vec C) 
&&  \opsimeq_{t \to + \infty}   
   e^{- t 2 e_1 }  P^{ini}(\vec C)    
\sum_{i=1}^N \sum_{c_i}  p_*(c_i)   \langle l_1^{[i]} \vert  P^{ini} \rangle
 l_1^2(c_i) \left[  \langle l_1^{[i]} \vert  P^{ini} \rangle -  l_1(C_i) \right]
    +...
 \nonumber \\
 &&    \opsimeq_{t \to + \infty}   
   e^{- t 2 e_1 } 
    P^{ini}(\vec C)   
\sum_{i=1}^N  \langle l_1^{[i]} \vert  P^{ini} \rangle \left[  \langle l_1^{[i]} \vert  P^{ini} \rangle -  l_1(C_i) \right]
    +...
\label{B0reconstructedbackwarddifferenceleading}
\end{eqnarray}


\subsection{ Convergence of observables in the backward reconstructive dynamics } 

Since the left eigenvectors $L_{\vec \gamma}(\vec C)$ form the basis of observables that display
the simple exponential behaviors of Eq. \ref{leftobservableGlobal} via the forward dynamics,
it is natural to analyze how these observables evolve via the reconstructive backward dynamics.

\subsubsection{ Analysis for a general left observable $L_{\vec \gamma \ne \vec 0}(\vec C) $}

To see how the observable $ L_{\vec \gamma \ne \vec 0}(\vec C)$   
converges in the reconstructive distribution $ B^{[0,t]}(\vec C) $
 towards its asymptotic value $L_{\vec \gamma}^{ini} = \langle L_{\vec \gamma} \vert  P^{ini} \rangle  $ in the initial distribution $P^{ini}(\vec C) $,
let us multiply the difference of Eq.  \ref{B0reconstructedbackwarddifference} by $L_{\vec \gamma}(\vec C) $
and sum over $\vec C$
  \begin{eqnarray}
\sum_{\vec C}  L_{\vec \gamma}(\vec C) \left[ B^{[0,t]}(\vec C) -  P^{ini}(\vec C) \right]
&&  =  \sum_{\vec c}   P_*(\vec c)
  \frac{ \displaystyle     \sum_{\vec \alpha \ne \vec 0} e^{- t E_{\vec \alpha} } 
L_{\vec \alpha}(\vec c) \sum_{\vec C}   \left[ L_{\vec \alpha}(\vec C) -  L_{\vec \alpha}^{ini}  \right] 
L_{\vec \gamma}(\vec C) P^{ini}(\vec C) }
 { \displaystyle 1 + \sum_{\vec \beta \ne \vec 0} e^{- t E_{\vec \beta} } L_{\vec \beta}(\vec c) 
 L_{\vec \beta}^{ini}   } 
 \nonumber \\
&&  =  \sum_{\vec c}   P_*(\vec c)
  \frac{ \displaystyle     \sum_{\vec \alpha \ne \vec 0} e^{- t E_{\vec \alpha} } 
L_{\vec \alpha}(\vec c)  
\bigg( \langle L_{\vec \alpha} L_{\vec \gamma} \vert  P^{ini} \rangle 
-   L_{\vec \alpha}^{ini}  L_{\vec \gamma}^{ini} \bigg)  }
 { \displaystyle 1 + \sum_{\vec \beta \ne \vec 0} e^{- t E_{\vec \beta} } L_{\vec \beta}(\vec c) 
 L_{\vec \beta}^{ini}  }   
\label{B0reconstructedbackwarddifferencebis}
\end{eqnarray}
The numerator thus involves the connected correlations that exist in the initial distribution $P^{ini} $
between the left observable $L_{\vec \gamma \ne \vec 0} $ under study and all the left observables $L_{\vec \alpha \ne \vec 0} $
 \begin{eqnarray}
&& \langle L_{\vec \alpha} L_{\vec \gamma} \vert  P^{ini} \rangle 
-  L_{\vec \alpha}^{ini}  L_{\vec \gamma}^{ini} 
=   \sum_{\vec C}     L_{\vec \alpha}(\vec C)  L_{\vec \gamma}(\vec C) P^{ini}(\vec C)  
-   \left[  \sum_{\vec C}     L_{\vec \alpha}(\vec C)   P^{ini}(\vec C)\right] 
 \left[  \sum_{\vec C'}      L_{\vec \gamma}(\vec C') P^{ini}(\vec C')\right] 
 \nonumber \\
 && =   \sum_{\vec C}    P^{ini}(\vec C)   \prod_{n=1}^N l_{\alpha_n}(C_n)l_{\gamma_n}(C_n)
-   \left[  \sum_{\vec C}       P^{ini}(\vec C) \prod_{n=1}^N l_{\alpha_n}(C_n)\right] 
 \left[  \sum_{\vec C'}       P^{ini}(\vec C')\prod_{m=1}^N l_{\gamma_m}(C_m')\right] 
 \label{ConnectedLeftIni}
\end{eqnarray}

To obtain the leading exponential behavior,
one can return to Eq \ref{B0reconstructedbackwarddifferenceleading}
in order to analyze the consequences for the observable $ L_{\vec \gamma \ne \vec 0}(\vec C)$
  \begin{eqnarray}
\sum_{\vec C}  L_{\vec \gamma}(\vec C) \left[ B^{[0,t]}(\vec C) -  P^{ini}(\vec C) \right]
&&    \opsimeq_{t \to + \infty}   
   e^{- t 2 e_1 } 
\sum_{i=1}^N  \langle l_1^{[i]} \vert  P^{ini} \rangle
 \sum_{\vec C}  \left[ \langle l_1^{[i]} \vert  P^{ini} \rangle -  l_1(C_i) \right] L_{\vec \gamma}(\vec C) P^{ini}(\vec C)  
    +...
    \nonumber \\
&& \opsimeq_{t \to + \infty}   
-   e^{- t 2 e_1 }  
\sum_{i=1}^N  \langle l_1^{[i]} \vert  P^{ini} \rangle
\left[  \langle l_1^{[i]} L_{\vec \gamma}  \vert  P^{ini} \rangle
 - \langle l_1^{[i]} \vert  P^{ini} \rangle \langle L_{\vec \gamma} \vert  P^{ini} \rangle 
   \right]  
    +...    
    \nonumber \\
\label{B0reconstructedbackwarddifferenceleadingobsleft}
\end{eqnarray}
Here the last factor corresponds to the connected correlations that exist in the initial distribution $P^{ini} $
between the left observable $L_{\vec \gamma \ne \vec 0} $ under study and
 the one-pixel-observables $ l_1^{[i]} $ associated to the $N$ pixels $i=1,..,N$
 \begin{eqnarray}
&& \langle l_1^{[i]} L_{\vec \gamma} \vert  P^{ini} \rangle 
-   \langle l_1^{[i]} \vert  P^{ini} \rangle \langle L_{\vec \gamma} \vert  P^{ini} \rangle 
=   \sum_{\vec C}     l_1(C_i)  L_{\vec \gamma}(\vec C) P^{ini}(\vec C)  
-   \left[  \sum_{\vec C}      l_1(C_i)  P^{ini}(\vec C)\right] 
 \left[  \sum_{\vec C'}      L_{\vec \gamma}(\vec C') P^{ini}(\vec C')\right] 
 \nonumber \\
 && =   \sum_{\vec C}    P^{ini}(\vec C)    l_1(C_i) \prod_{n=1}^N l_{\gamma_n}(C_n)
-   \left[  \sum_{\vec C}       P^{ini}(\vec C)  l_1(C_i)\right] 
 \left[  \sum_{\vec C'}       P^{ini}(\vec C')\prod_{m=1}^N l_{\gamma_m}(C_m')\right] 
 \label{ConnectedLeftInionepixel}
\end{eqnarray}


\subsubsection{ Special case of the one-pixel-observable $l_{\gamma \ne 0}(C_j) $ associated to the single pixel $j$}

For a single non-vanishing coefficient $\gamma_n=\gamma \delta_{n,j}$ corresponding 
to the observable $L_{\vec \gamma}(\vec C)=l_{\gamma}(C_j) $ associated to the single pixel $j$,
Eq. \ref{B0reconstructedbackwarddifferencebis} becomes
  \begin{eqnarray}
\sum_{\vec C}  l_{\gamma}(C_j) \left[ B^{[0,t]}(\vec C) -  P^{ini}(\vec C) \right]
 =  \sum_{\vec c}   P_*(\vec c)
  \frac{ \displaystyle     \sum_{\vec \alpha \ne \vec 0} e^{- t E_{\vec \alpha} } 
L_{\vec \alpha}(\vec c)  
\bigg( \langle L_{\vec \alpha} l_{\gamma}^{[j]} \vert  P^{ini} \rangle 
-   \langle L_{\vec \alpha} \vert  P^{ini} \rangle \langle  l_{\gamma}^{[j]}  \vert  P^{ini} \rangle \bigg)  }
 { \displaystyle 1 + \sum_{\vec \beta \ne \vec 0} e^{- t E_{\vec \beta} } L_{\vec \beta}(\vec c) 
 L_{\vec \beta}^{ini} \rangle  }   
\label{B0reconstructedbackwarddifferencebisl1gamma}
\end{eqnarray}
where the numerator involves the connected correlations that exist in the initial distribution $P^{ini} $
between the observable $l_{ \gamma }^{[j]} $ for the pixel $j$ and the observables $L_{\vec \alpha \ne \vec 0} $.

The asymptotic behavior of Eq. \ref{B0reconstructedbackwarddifferenceleadingobsleft} reduces to
  \begin{eqnarray}
\sum_{\vec C} l_{ \gamma }(C_j) \left[ B^{[0,t]}(\vec C) -  P^{ini}(\vec C) \right]
&&    \opsimeq_{t \to + \infty}   
-   e^{- t 2 e_1 }  
\sum_{i=1}^N  \langle l_1^{[i]} \vert  P^{ini} \rangle
\left[  \langle  l_1^{[i]}  l_{ \gamma }^{[j]}  \vert  P^{ini} \rangle
 - \langle l_1^{[i]} \vert  P^{ini} \rangle \langle l_{ \gamma }^{[j]} \vert  P^{ini} \rangle 
   \right]  
    +...    
    \nonumber \\
\label{B0reconstructedbackwarddifferenceleadingobsleftl1}
\end{eqnarray}
that involves the connected correlations that exist in the initial distribution $P^{ini} $
between the one-pixel left observable $l_{ \gamma }^{[j]}  $ for the pixel $j$ under study and
 the $N$ one-pixel left observables $ l_1^{[i]} $ with $i=1,..,N$
\begin{eqnarray}
&& \langle l_1^{[i]} l_{ \gamma }^{[j]} \vert  P^{ini} \rangle 
-   \langle l_1^{[i]} \vert  P^{ini} \rangle \langle l_{ \gamma }^{[j]} \vert  P^{ini} \rangle 
=   \sum_{\vec C}     l_1(C_i)   l_{ \gamma }(C_j) P^{ini}(\vec C)  
-   \left[  \sum_{\vec C}      l_1(C_i)  P^{ini}(\vec C)\right] 
 \left[  \sum_{\vec C'}       l_{ \gamma }(C_j') P^{ini}(\vec C')\right] 
 \label{ConnectedLeftInitwopixels}
\end{eqnarray}


\subsubsection{ Discussion }

In summary, the basis of observables associated to the left eigenvectors 
is very useful to analyze the convergence properties of the reconstructive backward dynamics.
This general framework is applied to spin models and to diffusion processes
in the sections \ref{sec_spins} and \ref{sec_diff} respectively.


\subsection{ Special case where the initial condition
is the empirical distribution constructed from $A$ images $\vec C^{[a]}$ }

When the initial condition $P^{ini}(\vec C)$
corresponds to the empirical distribution constructed from $A$ images $\vec C^{[a=1,2,..,A]}$
 \begin{eqnarray}
P^{ini}_{empi}( \vec C) = \frac{1}{A} \sum_{a=1}^A \delta_{\vec C,\vec C^{[a]}}
\label{empiC}
\end{eqnarray}
the backward propagator of Eq. \ref{backwardpropagator}
  \begin{eqnarray}
 B(  \vec C, 0 \vert \vec c, t ) 
&& =  \frac{ \langle \vec c \ \vert e^{  t W} \vert \vec C \rangle P^{ini}_{empi}(\vec C) }
 { \displaystyle \sum_{ \vec C'}  \langle \vec c \ \vert e^{  t W} \vert \vec C' \rangle P^{ini}_{empi}(\vec C')  }
 =\sum_{a=1}^A \delta_{\vec C,\vec C^{[a]}}  \frac{ \langle \vec c \ \vert e^{  t W} \vert \vec C^{[a]} \rangle  }
 { \displaystyle \sum_{a'=1}^A  \langle \vec c \ \vert e^{  t W} \vert \vec C^{[a']} \rangle   }
 \equiv \sum_{a=1}^A \delta_{\vec C,\vec C^{[a]}} \pi( \vec C^{[a]} \vert \vec c, t )
\label{backwardpropagatorempi}
\end{eqnarray}
is localized on the $A$ initial images $\vec C^{[a]} $, corresponding to the phenomenon of the perfect memorization
of the initial data that one actually wishes to avoid in practical applications (see \cite{Ambrogioni2,Biroli_dynamical,Ambrogioni5,forgery,generalization} and references therein),
while the weights $\pi( \vec C^{[a]} \vert \vec c, t ) $ of the $A$ images depend on the time $t$ and on the  configuration $\vec c$ via the $A$ forward propagators $\langle \vec c \ \vert e^{  t W} \vert \vec C^{[a]} \rangle $
  \begin{eqnarray}
\pi( \vec C^{[a]} \vert \vec c, t ) \equiv  \frac{ \langle \vec c \ \vert e^{  t W} \vert \vec C^{[a]} \rangle  }
 { \displaystyle \sum_{a'=1}^A  \langle \vec c \ \vert e^{  t W} \vert \vec C^{[a']} \rangle   }
\label{backwardpropagatorempiwadef}
\end{eqnarray}

The asymptotic behavior of Eq. \ref{B0reconstructedbackwarddifferenceleading}
 \begin{eqnarray}
 B^{[0,t]}(\vec C) 
 &&    \opsimeq_{t \to + \infty}   
  P^{ini}_{empi}(\vec C)   \left( 1+    e^{- t 2 e_1 } 
   \sum_{i=1}^N  \langle l_1^{[i]} \vert  P^{ini} \rangle \left[  \langle l_1^{[i]} \vert  P^{ini} \rangle -  l_1(C_i) \right]
       \right)+...
   \nonumber \\
   &&
  \opsimeq_{t \to + \infty}   
 \sum_{a=1}^A \delta_{\vec C,\vec C^{[a]}}   \frac{1}{A}  \left( 1+    e^{- t 2 e_1 } 
   \sum_{i=1}^N  \left[\frac{1}{A} \sum_{a'=1}^A  l_1(C_i^{[a']})  \right] \left[  \frac{1}{A} \sum_{a''=1}^A  l_1(C_i^{[a'']})  -  l_1(C_i^{[a]}) \right]
       \right)+...
\label{B0reconstructedbackwarddifferenceleadingempi}
\end{eqnarray}
then yields that the convergence of the weights $\pi( \vec C^{[a]} \vert \vec c, t ) $ of the $A$ images $\vec C^{[a]} $ towards their equiprobable value $\frac{1}{A}$ 
  \begin{eqnarray}
\pi( \vec C^{[a]} \vert \vec c, t )   \opsimeq_{t \to + \infty} 
 \frac{1}{A}  \left( 1+    e^{- t 2 e_1 } 
   \sum_{i=1}^N  \left[\frac{1}{A} \sum_{a'=1}^A  l_1(C_i^{[a']})  \right] \left[  \frac{1}{A} \sum_{a''=1}^A  l_1(C_i^{[a'']})  -  l_1(C_i^{[a]}) \right]
       \right)+...
\label{backwardpropagatorempiwa}
\end{eqnarray}
involves the exponential factor $e^{- t 2 e_1 } $ and  the values of the first left eigenvector $l_1(C_i^{[a']}) $ for 
the $N$ pixels $i=1,..,N$ of the $A$ images $a'=1,..,A$.


\section{ Application to single-spin-flip Markov generative models  }

\label{sec_spins}

In this section, we focus on the case where each pixel $n=1,..,N$ can only be in the two configurations $S_i=\pm$,
so that there are $2^N$ possible global configurations $\vec S=(S_1,..,S_N)$ for the $N$ pixels.
The forwards dynamics is the independent single-spin-flip dynamics considered in \cite{garrahan}.
Other examples of generative models based on Markov chains on discrete-state spaces can be found in \cite{Discrete_stuctured,Discrete_denoising,Discrete_markov,Discrete_score,Discrete_guidance} and references therein.

\subsection{ Spectral properties of the forward dynamics for a single pixel }

The simplest continuous-time Markov jump dynamics for the probabilities $p_t(S)$ of the two configurations
 $S=\pm 1$ of a single pixel involves some flip rate $\zeta>0$ 
 \begin{eqnarray}
\partial_t p_t(+) && =  - \zeta  p_t(+)+\zeta  p_t(-)
\nonumber \\
\partial_t p_t(-) && = \zeta  p_t(+) - \zeta  p_t(-) 
\label{spinflip}
\end{eqnarray}
and converges towards the steady state where the two values $S=\pm$ have the same probabilities
 \begin{eqnarray}
p_*(S=\pm) = \frac{1}{2}
\label{spinflipsteady}
\end{eqnarray}
As a consequence, the similarity transformation of Eq. \ref{similarity} is trivial,
and the $2 \times 2$ Markov matrix $w$ governing the dynamics of Eq. \ref{spinflip}
 \begin{eqnarray}
w = \begin{pmatrix} 
- \zeta &   \zeta \\
 \zeta & - \zeta 
 \end{pmatrix} = \zeta (\sigma^x - \mathbb{1} ) = - H
\label{wflip}
\end{eqnarray}
can be directly interpreted as the opposite of a quantum Hamiltonian $H$
that involves the identity $\mathbb{1} $
and the Pauli matrix $\sigma^x $.
The spectral decomposition of $\sigma^x $ involving its two eigenvalues $(\pm)$ 
 \begin{eqnarray}
\sigma^x = \begin{pmatrix} 
0 &   1 \\
1 & 0 
 \end{pmatrix} = \vert \sigma^x=+1 \rangle \langle  \sigma^x=+1 \vert 
- \vert \sigma^x=-1 \rangle \langle  \sigma^x=-1 \vert 
\label{paulix}
\end{eqnarray}
yields that the spectral decomposition of the evolution operator $e^{-t H}$
  \begin{eqnarray}
 e^{ - t H} =  e^{  t \zeta (\sigma^x - \mathbb{1} )} 
&& =  \vert \sigma^x=+1 \rangle \langle  \sigma^x=+1 \vert 
+ e^{-t 2 \zeta}  \vert \sigma^x=-1 \rangle \langle  \sigma^x=-1 \vert
\nonumber \\
&& \equiv   \vert \psi_0 \rangle \langle  \psi_0 \vert 
+ e^{-t e_1 } \vert \psi_1 \rangle \langle  \psi_1 \vert 
\label{spectralHspins}
\end{eqnarray}
can be identified with the notations of Eq. \ref{spectralH} with its two terms $\alpha=0,1$ as follows :
 the groundstate $\vert \psi_0 \rangle $ associated to the vanishing energy $e_0=0$
and the excited state $\vert \psi_1 \rangle $ associated to the energy 
  \begin{eqnarray}
 e_1 = 2 \zeta
\label{e1spin}
\end{eqnarray}
are simply the two eigenstates of the Pauli matrix $\sigma^x$
  \begin{eqnarray}
\vert \psi_0 \rangle && =\vert \sigma^x=+1 \rangle = \frac{\vert + \rangle +\vert - \rangle }{\sqrt 2}
\nonumber \\
\vert \psi_1 \rangle && =\vert \sigma^x=-1 \rangle = \frac{\vert + \rangle -\vert - \rangle }{\sqrt 2}
\label{paulisigmaxeigen}
\end{eqnarray}
The translation into the left and right eigenvectors via Eq. \ref{eigenwpsi}
is in agreement with Eq. \ref{eigenw0}
for $\alpha=0$ that involves the steady state of Eq. \ref{spinflipsteady}
  \begin{eqnarray}
 \langle  l_0 \vert && =  \sqrt{2}  \langle  \psi_0 \vert  =   \langle  + \vert +  \langle  - \vert
\nonumber \\
  \vert  r_0 \rangle && =  \frac{1}{ \sqrt{2} } \vert \psi_0 \rangle  = \frac{\vert + \rangle +\vert - \rangle }{ 2} = \vert  p_* \rangle
\label{eigenw0spin}
\end{eqnarray}
and yields for $\alpha=1$
  \begin{eqnarray}
 \langle  l_1 \vert && =  \sqrt{2}  \langle  \psi_1 \vert = \langle  + \vert -  \langle  - \vert
\nonumber \\
  \vert  r_1 \rangle && =  \frac{1}{ \sqrt{2} } \vert \psi_1 \rangle = \frac{\vert + \rangle -\vert - \rangle }{ 2}
\label{eigenw1spin}
\end{eqnarray}
As a consequence, the observable $l_1^{av}(t)  $  of Eq. \ref{leftobservable}
associated to the excited left eigenvector 
  \begin{eqnarray}
 l_1(S) = S
\label{l1spin}
\end{eqnarray}
simply represents the magnetization ${\cal M}(t)$ at time $t$
  \begin{eqnarray}
 l_1^{av}(t) \equiv \sum_{S=\pm} l_1(S) p_t( S) = \sum_{S=\pm} S p_t( S)= p_t(+)-p_t(-) \equiv {\cal M}(t)
\label{l1observable}
\end{eqnarray}
and displays the exponential decay involving the excited energy $e_1$ 
  \begin{eqnarray}
 {\cal M}(t)  = e^{-t e_1 }   {\cal M}(0) 
\label{l1decay}
\end{eqnarray}

The spectral decomposition of Eq. \ref{spectralpixel} for
the finite-time propagator reduces to
  \begin{eqnarray}
 \langle s \vert e^{  t w} \vert S \rangle 
 = p_*(s)+ e^{-t e_1 }r_1 (s) l_1(S)
 = \frac{1}{2} + e^{-t e_1 }  \frac{s S}{2}
\label{spectral1spin}
\end{eqnarray}
and can be rewritten in various analytical forms since the two spins can only take the two values $(\pm 1)$ :
for instance one can choose to distinguish the two cases $s=\pm S$ with delta functions
in order to make the link with an expression given in \cite{garrahan}
  \begin{eqnarray}
 \langle s \vert e^{  t w} \vert S \rangle = \frac{ 1+ e^{-t e_1} }{2} \delta_{s,S}
+ \frac{ 1- e^{-t e_1} }{2} \delta_{s,-S}
\label{spectral1spindelta}
\end{eqnarray}
or one can put the product $(sS)=\pm 1$ in the exponential via
 \begin{eqnarray}
 \langle s \vert e^{  t w} \vert S \rangle 
 = \frac{\sqrt{(1+ e^{-t e_1})(1- e^{-t e_1})}}{2} \left( \sqrt{ \frac{(1+ e^{-t e_1})}{(1- e^{-t e_1})}}\right)^{sS}
 = \frac{\sqrt{1- e^{-t 2 e_1}}}{2} \left( \sqrt{ \frac{1+ e^{-t e_1}}{1- e^{-t e_1}}}\right)^{sS}
\label{spectral1spinexpo}
\end{eqnarray}

This finite-time propagator $\langle s \vert e^{  t w} \vert S \rangle $ concerning
 the continuous-time Markov jump dynamics for the spins
has actually the same form as the propagator associated to the discrete-time Markov chain
considered in \cite{c_generativeLargedevPixels}
 up to the correspondance of notations $\lambda = e^{-e_1 }$.
 As a consequence, all the results that are based on this explicit finite-time propagator
can be directly translated between the continuous-time and discrete-time frameworks.


\subsection{ Spectral properties of the forward dynamics for the $N$ pixels }

\subsubsection{ Interpretation of the spectrum in terms of the relaxation of the initial spin-correlations }

The $2^N$ energies $E_{\vec \alpha=(\alpha_1,..,\alpha_N)} $ of Eq. \ref{Esum}
  \begin{eqnarray}
E_{\vec \alpha} = \sum_{n=1}^N e_{\alpha_n} = e_1 \sum_{n=1}^N  \delta_{\alpha_n,1} 
\label{EsumSpins}
\end{eqnarray}
simply counts the number $K \in \{0,1,2,..,N\}$ of non-vanishing indices $\alpha_n=1$ among
the $N$ indices $(\alpha_1,..,\alpha_N)$.
The vanishing global energy $E_{\vec \alpha=\vec 0} =0$ is associated
to the global steady state 
  \begin{eqnarray}
P_*( \vec s) =  \prod_{n=1}^N p_*(s_n) = \frac{1}{2^N}
\label{SteadyGlobalSpins}
\end{eqnarray}
where the $2^N$ configurations of the $N$ spins have the same probability $\frac{1}{2^N} $.

Using $l_0$ of Eq. \ref{eigenw0spin} and $l_1$ of \ref{l1spin}, one obtains that
the global left eigenvectors of Eq. \ref{Lprod}  reduce to
  \begin{eqnarray}
  L_{\vec \alpha} (\vec S) && = \prod_{n=1}^N l_{\alpha_n}(S_n) = \prod_{n=1}^N S_n^{\alpha_n}
\label{LeftSpins}
\end{eqnarray}
As a consequence, the $(2^N-1)$ observables $L^{av}_{\vec \alpha \ne \vec 0} (t) $ of Eq. \ref{leftobservable}
  \begin{eqnarray}
  L^{av}_{\vec \alpha} (t)  \equiv  \sum_{\vec S}   L_{\vec \alpha} (\vec S) P(\vec S,t) = 
  \sum_{\vec S}   \left[ \prod_{n=1}^N S_n^{\alpha_n} \right] P(\vec S,t) 
  \label{Lavspin}
\end{eqnarray}
correspond to all the correlation functions that can be constructed out of the $N$ spins
and display the simple exponential decay
  \begin{eqnarray}
   L^{av}_{\vec \alpha} (t) = e^{-t E_{\vec \alpha} }  L^{av}_{\vec \alpha} (t=0) = e^{-t E_{\vec \alpha} }  L^{ini}_{\vec \alpha}
  \label{Corredyn}
\end{eqnarray}
starting from their values of Eq. \ref{Leftini}  in the initial distribution $P^{ini}(\vec S) $ 
 \begin{eqnarray}
 L_{\vec \alpha}^{ini} =   \sum_{\vec S}    \left[ \prod_{n=1}^N S_n^{\alpha_n} \right]  P^{ini}(\vec S)  
  \label{correini}
\end{eqnarray}

In particular, the first excited energy $e_1$ of Eq. \ref{Efirstexcited}
is $N$ times degenerate and governs the relaxations toward zero of the $N$ individual magnetizations ${\cal M}_i(t)$
discussed in Eqs \ref{l1observable}
and \ref{l1decay} 
  \begin{eqnarray}
  L^{av}_{(\alpha_n=\delta_{n,i})} (t)  \equiv \sum_{\vec S} S_{i}  P(\vec S,t) \equiv  {\cal M}_i(t) = e^{- t e_1} {\cal M}_i^{ini} 
  \label{LfirstExcitedspin}
\end{eqnarray}
The second global excited energy $[2 e_1]$ is associated to the relaxation of the $\frac{N(N-1)}{2}$ correlations 
${\cal C}_{i_1,i_2}(t) $ between two spins $S_{i_1} $ and $S_{i_2} $ with $1 \leq i_1 < i_2 \leq N$
  \begin{eqnarray}
  L^{av}_{(\alpha_n=\delta_{n,i_1}+\delta_{n,i_2})} (t)  \equiv  \sum_{\vec S} S_{i_1} S_{i_2} P(\vec S,t)
  \equiv {\cal C}_{i_1,i_2}(t) 
  =  e^{- t 2 e_1} {\cal C}_{i_1,i_2}^{ini}
  \label{LsecondExcitedspinobs}
\end{eqnarray}
More generally, the $K^{th}$ global excited energy $[K e_1]$ with $K \in \{1,2,..,N\}$ governs the relaxations of the $ \frac{N!}{K! (N-K)!} $ correlations $ {\cal C}_{i_1,i_2,..,i_K}$
between the $K$ spins $1 \leq i_1<i_2<.. < i_K \leq N$
  \begin{eqnarray}
  L^{av}_{(\alpha_n=\sum_{k=1}^K \delta_{n,i_k})} (t)  
  \equiv  \sum_{\vec S} \left[\prod_{k=1}^K S_{i_k} \right] P(\vec S,t) \equiv {\cal C}_{i_1,i_2,..,i_K}(t)
  =  e^{- t K e_1} {\cal C}_{i_1,i_2,..,i_K}^{ini}
  \label{LKExcitedspinobs}
\end{eqnarray}


\subsubsection{ Spectral decomposition of the global forward propagator }

The global forward propagator of Eq. \ref{factorizedpropagatorspectral}
reduces to the product of the $N$ individual propagators of Eq. \ref{spectral1spin} or \ref{spectral1spinexpo}
  \begin{eqnarray}
  \langle \vec s \vert e^{  t W} \vert \vec S \rangle
= \prod_{n=1}^N  \langle s_n \vert e^{  t w_n} \vert S_n \rangle
&&  = \prod_{n=1}^N \left[ \frac{1 +  e^{-t e_1 }  s_n S_n}{2} \right]
\nonumber \\
&& = \left( \frac{\sqrt{1- e^{-t 2 e_1}}}{2}\right)^N 
 \left( \sqrt{ \frac{1+ e^{-t e_1}}{1- e^{-t e_1}}}\right)^{ \sum_{n=1}^N s_nS_n}
\label{factorizedpropagatorspectralspins}
\end{eqnarray}
This form of the global forward propagator suggests to analyze its large deviations properties with respect to the number $N$ of pixels as described in detail in \cite{c_generativeLargedevPixels},
while here we will focus on its spectral decomposition
of Eq. \ref{factorizedpropagatorspectral}
in terms of the left eigenvectors of Eq. \ref{LeftSpins}
  \begin{eqnarray}
&& \langle \vec s \vert e^{  t W} \vert \vec S \rangle 
  = P_*( \vec s) \bigg[ 1+ \sum_{\vec \alpha \ne \vec 0} e^{- t E_{\vec \alpha} } L_{\vec \alpha}(\vec s) L_{\vec \alpha}(\vec S) \bigg]
\nonumber \\
&& =  \frac{1}{2^N} \bigg[ 1 +  e^{- t e_1 } \sum_{i=1}^N s_i S_i
  +  e^{- t 2 e_1 } \sum_{1 \leq i_1 < i_2 \leq N} s_{i_1} s_{i_2} S_{i_1} S_{i_2}
  + \sum_{K=3}^N e^{- t K e_1 } \sum_{1 \leq i_1 < i_2 ..<i_K \leq N} s_{i_1} s_{i_2}..s_{i_K} S_{i_1} S_{i_2}..S_{i_K}
   \bigg]
   \ \ \ 
\label{factorizedpropagatorspectralonlyleftspins}
\end{eqnarray}

The application of the forward propagator 
to the initial distribution $P^{ini}(\vec S) $ yields the probability distribution at time $t$ of Eq. \ref{ptforward}
in terms of the initial spin-correlations
$ L_{\vec \alpha}^{ini} $ of Eq. \ref{correini}
 \begin{eqnarray}
p(\vec s,t) && = \sum_{\vec S}  \langle \vec s \vert e^{  t W} \vert \vec S \rangle P^{ini}(\vec S)
 =    \sum_{\vec S}  \frac{1}{2^N} \bigg[ 1+ \sum_{\vec \alpha \ne \vec 0} 
  \prod_{n=1}^N \left( e^{-t e_1 }  s_n S_n\right)^{\alpha_n} \bigg] P^{ini}(\vec S)
  =     \frac{1}{2^N} \left( 1 + \sum_{\vec \alpha \ne \vec 0}   L^{ini}_{\vec \alpha}
 \left[ \prod_{n=1}^N \left( e^{-t e_1} s_n\right)^{\alpha_n} \right] \right)
 \nonumber \\
&& 
  =   \frac{1}{2^N} \left( 1 +  e^{- t e_1 } \sum_{i=1}^N s_i {\cal M}^{ini}_i
  +  e^{- t 2 e_1 } \sum_{1 \leq i_1 < i_2 \leq N} s_{i_1} s_{i_2}{\cal C}^{ini}_{i_1,i_2}
  + \sum_{K=3}^N e^{- t K e_1 } \sum_{1 \leq i_1 < i_2 ..<i_K \leq N} s_{i_1} s_{i_2}..s_{i_K}{\cal C}^{ini}_{i_1,i_2,..,i_K}
   \right)\ \ \ \ 
\label{Probafromcorredynexpli}
\end{eqnarray}


\subsubsection{ Convergence properties of the forwards dynamics }

As discussed around Eq. \ref{factorizedpropagatorspectralonlyleftfirstexcited},
 the convergence of the propagator of Eq. \ref{factorizedpropagatorspectralonlyleftspins} towards the steady state $P_*(\vec s)=\frac{1}{2^N} $
is governed by the first excited energy $e_1$
  \begin{eqnarray}
P( \vec s, t\vert \vec S, 0) 
 =  \frac{1}{2^N} \bigg[ 1+  e^{- t e_1 } \sum_{i=1}^N s_i S_i +...\bigg]
\label{factorizedpropagatorspectralonlyleftfirstexcitedspins}
\end{eqnarray}

The Kemeny-time of Eq. \ref{tauspectral} can be evaluated either from the direct sum over the $N$ excited energies
of Eq. \ref{EsumSpins} with their binomial degeneracies
 \begin{eqnarray}
\tau^{Spectral}_N 
\equiv  \sum_{\vec \alpha \ne \vec 0} \frac{1}{E_{\vec \alpha}} = \sum_{K=1}^N \frac{1}{K e_1} \times \frac{ N! }{  K! (N-K)! } 
\label{tauspectralsumspin}
\end{eqnarray}
or via the integral of Eq. \ref{tauspectralN} that involves the single-pixel spectral partition function of Eq. \ref{SpectralPartitionz1} that reduces to $Z_1(t) =   e^{- t e_1} $
\begin{eqnarray}
\tau^{Spectral}_N =    \int_0^{+\infty} dt  \left(  \left[ 1+Z_1(t) \right]^N -1\right) 
=  \int_0^{+\infty} dt  \left[  \left( 1+e^{- t e_1}  \right)^N -1\right]
= \frac{1}{e_1} \int_0^1 \frac{du}{u}  \left[  ( 1+u )^N -1\right]
\label{tauspectralNspins}
\end{eqnarray}
displaying the exponential growth of Eq. \ref{tauspectralNlarge}
with $\alpha_{max}=1$
 \begin{eqnarray}
\tau^{Spectral}_N \opsimeq_{N \to + \infty}   \frac{2^{N+1}}{  N e_1 }
\label{tauspectralNlargespins}
\end{eqnarray}

The Mean-First-Passage-Time $\tau(\vec s \vert \vec S)$ of Eq. \ref{MFTPspectralleft}
reads using the left eigenvectors of Eq. \ref{LeftSpins}
 \begin{eqnarray}
   \tau(\vec s \vert \vec S) && =  \sum_{\vec \alpha \ne \vec 0}  \frac{ L_{\vec \alpha} (\vec s)  \left[ L_{\vec \alpha} (\vec s) - L_{\vec \alpha} (\vec S)\right] }{E_{\vec \alpha}}
      \nonumber \\
   && = \sum_{K=1}^N \frac{1}{K e_1} \sum_{1 \leq i_1<i_2<..<i_K \leq N} 
  \left[ 1 - \prod_{k=1}^K s_{i_k} S_{i_k} \right]
\label{MFTPspectralleftspins}
\end{eqnarray}


\subsection{ Properties of the reconstructive backward dynamics }

The backward propagator of Eqs \ref{backwardpropagator} and \ref{backwardpropagatorspectral}
reads using Eq. \ref{factorizedpropagatorspectralonlyleftspins}
  \begin{eqnarray}
&& B(  \vec S, 0 \vert \vec s, t ) = \frac{ \langle \vec s \ \vert e^{  t W} \vert \vec S \rangle P^{ini}(\vec S) }{  P( \vec s, t) }
\nonumber \\
&&  = P^{ini}(\vec S)
   \frac{ \displaystyle  1 +  e^{- t e_1 } \sum_{i=1}^N s_i S_i
  +  e^{- t 2 e_1 } \sum_{1 \leq i_1 < i_2 \leq N} s_{i_1} s_{i_2} S_{i_1} S_{i_2}
  + \sum_{K=3}^N e^{- t K e_1 } \sum_{1 \leq i_1 < i_2 ..<i_K \leq N} s_{i_1} s_{i_2}..s_{i_K} S_{i_1} S_{i_2}..S_{i_K}   }
 { \displaystyle   1 +  e^{- t e_1 } \sum_{j=1}^N s_j {\cal M}^{ini}_j
  +  e^{- t 2 e_1 } \sum_{1 \leq j_1 < j_2 \leq N} s_{j_1} s_{j_2}{\cal C}^{ini}_{j_1,j_2}
  + \sum_{K=3}^N e^{- t K e_1 } \sum_{1 \leq j_1 < j_2 ..<j_K \leq N} s_{j_1} s_{j_2}..s_{j_K}{\cal C}^{ini}_{j_1,j_2,..,j_K}  }
  \ \ \ \ 
\label{backwardpropagatorspins}
\end{eqnarray}

The application of this finite-time backward-propagator 
to the steady state $P_*(\vec s)= \frac{1}{2^N}$ 
leads to the reconstructed distribution of Eq. \ref{B0reconstructedbackward}
  \begin{eqnarray}
&& B^{[0,t]}(\vec S)  = \sum_{\vec s}  B(  \vec S, 0 \vert \vec s, t )  \frac{1}{2^N}
\nonumber \\
&& = \frac{P^{ini}(\vec S)}{2^N} \sum_{\vec s}  
   \frac{ \displaystyle  1 +  e^{- t e_1 } \sum_{i=1}^N s_i S_i
  +  e^{- t 2 e_1 } \sum_{1 \leq i_1 < i_2 \leq N} s_{i_1} s_{i_2} S_{i_1} S_{i_2}
  + \sum_{K=3}^N e^{- t K e_1 } \sum_{1 \leq i_1 < i_2 ..<i_K \leq N} s_{i_1} s_{i_2}..s_{i_K} S_{i_1} S_{i_2}..S_{i_K}   }
 { \displaystyle   1 +  e^{- t e_1 } \sum_{j=1}^N s_j {\cal M}^{ini}_j
  +  e^{- t 2 e_1 } \sum_{1 \leq j_1 < j_2 \leq N} s_{j_1} s_{j_2}{\cal C}^{ini}_{j_1,j_2}
  + \sum_{K=3}^N e^{- t K e_1 } \sum_{1 \leq j_1 < j_2 ..<j_K \leq N} s_{j_1} s_{j_2}..s_{j_K}{\cal C}^{ini}_{j_1,j_2,..,j_K}  }
  \ \ \ \ 
\label{B0reconstructedbackwardspins}
\end{eqnarray}
so that the difference of Eq. \ref{B0reconstructedbackwarddifference}
with $P^{ini}(\vec S)  $ becomes 
\begin{small}
  \begin{eqnarray}
&& B^{[0,t]}(\vec S) -  P^{ini}(\vec S) 
 = \frac{P^{ini}(\vec S)}{2^N} 
\nonumber \\
&& \times \sum_{\vec s}      \frac{ \displaystyle   e^{- t e_1 } \sum_{i=1}^N s_i (S_i - {\cal M}^{ini}_i)
  +  e^{- t 2 e_1 } \sum_{1 \leq i_1 < i_2 \leq N} s_{i_1} s_{i_2} (S_{i_1} S_{i_2} - {\cal C}^{ini}_{i_1,i_2})
  + \sum_{K=3}^N e^{- t K e_1 } \sum_{1 \leq i_1 < i_2 ..<i_K \leq N} s_{i_1} s_{i_2}..s_{i_K} 
  (S_{i_1} S_{i_2}..S_{i_K} - {\cal C}^{ini}_{i_1,i_2,..,i_K})   }
 { \displaystyle   1 +  e^{- t e_1 } \sum_{j=1}^N s_j {\cal M}^{ini}_j
  +  e^{- t 2 e_1 } \sum_{1 \leq j_1 < j_2 \leq N} s_{j_1} s_{j_2}{\cal C}^{ini}_{j_1,j_2}
  + \sum_{K=3}^N e^{- t K e_1 } \sum_{1 \leq j_1 < j_2 ..<j_K \leq N} s_{j_1} s_{j_2}..s_{j_K}{\cal C}^{ini}_{j_1,j_2,..,j_K}  }
\nonumber \\
\label{B0reconstructedbackwarddifferencespins}
\end{eqnarray}
\end{small}


\subsubsection{ Leading asymptotic exponential convergence of $ B^{[0,t]}(\vec C)$ towards $P^{ini}(\vec C) $ }

In terms of the left eigenvectors $l_1(S_i)=S_i$
and of the magnetizations ${\cal M}_i^{ini}$ of the $N$ pixels $i=1,..,N$ in the initial distribution $P^{ini}$
 \begin{eqnarray}
{\cal M}_i^{ini} = \sum_{\vec S} S_{i}  P^{ini}(\vec S) = \langle l_1^{[i]} \vert  P^{ini} \rangle
  \label{Mini}
\end{eqnarray}
the asymptotic behavior of Eq. \ref{B0reconstructedbackwarddifferenceleading}
for the difference of Eq. \ref{B0reconstructedbackwarddifferencespins}
reads
  \begin{eqnarray}
 B^{[0,t]}(\vec S) -  P^{ini}(\vec S) 
 && \opsimeq_{t \to + \infty}   
    e^{- t 2 e_1 } 
    P^{ini}(\vec S)   
\sum_{i=1}^N  \langle l_1^{[i]} \vert  P^{ini} \rangle \left[  \langle l_1^{[i]} \vert  P^{ini} \rangle -  l_1(S_i) \right]
    +...
 \nonumber \\
&&     \opsimeq_{t \to + \infty}   
   e^{- t 2 e_1 }  P^{ini}(\vec S)   
\sum_{i=1}^N  {\cal M}_i^{ini}  \left[  {\cal M}_i^{ini} -  S_i \right]
    +...
\label{B0reconstructedbackwarddifferenceleadingspins}
\end{eqnarray}


\subsubsection{ Convergence of the correlations between the $K$ spins $1 \leq i_1<i_2<.. < i_K \leq N$}

For the left observable associated to the correlation between $K$ spins $1 \leq i_1<i_2<.. < i_K \leq N$
  \begin{eqnarray}
  L_{(\gamma_n=\sum_{k=1}^K \delta_{n,i_k})}     = \prod_{k=1}^K S_{i_k} 
  \label{LKobs}
\end{eqnarray}
the leading exponential behavior of Eq. \ref{B0reconstructedbackwarddifferenceleadingobsleft}
becomes
  \begin{eqnarray}
&& \sum_{\vec S}  \left[ \prod_{k=1}^K S_{i_k} \right] \left[ B^{[0,t]}(\vec S) -  P^{ini}(\vec S) \right]
 \opsimeq_{t \to + \infty}   
-   e^{- t 2 e_1 }  
\sum_{i=1}^N  {\cal M}^{ini}_i 
\left[  \sum_{\vec S}  P^{ini} (\vec S) S_i \prod_{k=1}^K S_{i_k} 
 - {\cal M}^{ini}_i { \cal C}^{ini}_{i_1,i_2,..,i_K}  
   \right]  
    +...    
    \nonumber \\
&& \opsimeq_{t \to + \infty}   
-   e^{- t 2 e_1 }  
\bigg( \sum_{l=1}^K  {\cal M}^{ini}_{i_l} 
\left[  { \cal C}^{ini}_{i_1,.,i_{l-1},i_{l+1}.,i_K}
 - {\cal M}^{ini}_{i_l} { \cal C}^{ini}_{i_1,i_2,..,i_K}  
   \right]  
 + 
\sum_{i \ne(i_1,..,i_K) }  {\cal M}^{ini}_i 
\left[   { \cal C}^{ini}_{i,i_1,i_2,..,i_K}
 - {\cal M}^{ini}_i { \cal C}^{ini}_{i_1,i_2,..,i_K} 
   \right]  \bigg)
    +...       
\label{B0reconstructedbackwarddifferenceleadingobsleftspins}
\end{eqnarray}
that involves the $N$ initial magnetizations ${\cal M}^{ini}_i $, 
the initial correlation ${ \cal C}^{ini}_{i_1,..,i_K} $ between the $K$ spins $i_{k=1,..,K}$,
the initial correlations ${ \cal C}^{ini}_{i_1,.,i_{l-1},i_{l+1}.,i_K} $
between the $(K-1)$ spins $i_k \ne i_l$ 
as well as the initial correlation ${ \cal C}^{ini}_{i,i_1,i_2,..,i_K} $
between the $(K+1)$ spins $i$ and $i_{k=1,..,K} $.

In particular, the convergence for the correlation between the $K=2$ spins $S_1$ and $S_2$
  \begin{eqnarray}
&& \sum_{\vec S}  S_1 S_2  B^{[0,t]}(\vec S) -  { \cal C}^{ini}_{1,2}
 \opsimeq_{t \to + \infty}   
 \nonumber \\
 && 
-   e^{- t 2 e_1 }  \left(
  {\cal M}^{ini}_1 
\left[  {\cal M}^{ini}_2
 - {\cal M}^{ini}_1 { \cal C}^{ini}_{1,2}  
   \right]  
    + {\cal M}^{ini}_2 
\left[  {\cal M}^{ini}_1
 - {\cal M}^{ini}_2 { \cal C}^{ini}_{1,2}  
   \right] 
   + \sum_{i =3 }^N  {\cal M}^{ini}_i 
\left[   { \cal C}^{ini}_{1,2,i}
 -  { \cal C}^{ini}_{1,2} {\cal M}^{ini}_i
   \right]  \right)     +...       
\label{B0reconstructedbackwarddifferenceleadingobsleftspins12}
\end{eqnarray}
involves the $N$ magnetizations ${\cal M}^{ini}_i $, the two-spin correlation ${ \cal C}^{ini}_{1,2} $
as well as the $(N-2)$ three-spin correlations ${ \cal C}^{ini}_{1,2,i} $ with $i=3,..,N$


\subsubsection{ Convergence of the magnetization of the spin $j$}

For a single non-vanishing coefficient $\gamma_n=\gamma \delta_{n,j}$ corresponding 
to the magnetization of the single pixel $j$,
the asymptotic behavior of Eq. \ref{B0reconstructedbackwarddifferenceleadingobsleftspins}
reduces to
  \begin{eqnarray}
\sum_{\vec S}  S_j B^{[0,t]}(\vec S) -  {\cal M}^{ini}_j 
&&    \opsimeq_{t \to + \infty}   
-   e^{- t 2 e_1 }   
\sum_{i=1}^N  {\cal M}^{ini}_i
\left[  \sum_{\vec S} S_i S_j P^{ini} (\vec S) 
 - {\cal M}^{ini}_i {\cal M}^{ini}_j 
   \right]  
    +...    
        \nonumber \\
&& \opsimeq_{t \to + \infty}   
-   e^{- t 2 e_1 }  \left(  {\cal M}^{ini}_j 
\left[  1 - ({\cal M}^{ini}_j )^2  
   \right]  
+
\sum_{i \ne j }  {\cal M}^{ini}_i 
\left[   { \cal C}^{ini}_{i,j} - {\cal M}^{ini}_i  {\cal M}^{ini}_j  
   \right]  
   \right)
    +...      
\label{B0reconstructedbackwarddifferenceleadingobsleftl1spins}
\end{eqnarray}
that involves the magnetizations ${\cal M}^{ini}_i $ of the $N$ spins
as well as the $(N-1)$ connected correlations between the spin $j$ and the other $(N-1)$ spins $i \ne j$.


\subsubsection{ Simple example with only $N=2$ pixels }

To see more concretely how the reconstruction works for any finite time $t$, 
let us write explicitly the full reconstructed distribution of Eq. \ref{B0reconstructedbackwardspins}
for the simple example involving only $N=2$ spins
  \begin{eqnarray}
B^{[0,t]}(S_1,S_2) && 
= \frac{P^{ini}(S_1,S_2)}{4} \sum_{s_1=\pm}  \sum_{s_2=\pm} 
\frac{ \displaystyle  1+  e^{-t e_1 }  (s_1 S_1 +s_2 S_2)  
+    e^{-t 2 e_1 }  s_1 s_2 S_1  S_2 }
 { \displaystyle   1 + e^{-t e_1} ( s_1 {\cal M}^{ini}_1 +s_2 {\cal M}^{ini}_2 ) 
   + e^{-t 2 e_1 }  s_1  s_2 { \cal C}^{ini}_{1,2}  }
\label{B0reconstructedbackwardspinsS1S2}
\end{eqnarray}
as well as the corresponding magnetizations of the two spins
  \begin{eqnarray}
\sum_{S_1=\pm 1} \sum_{S_2=\pm 1} S_1 B^{[0,t]}(S_1,S_2)    
&& =
 \frac{1}{4}  \sum_{s_1=\pm}  \sum_{s_2=\pm} 
\frac{ \displaystyle   {\cal M}^{ini}_1+  e^{-t e_1 }  (s_1  +s_2 { \cal C}^{ini}_{1,2})  
+    e^{-t 2 e_1 }  s_1 s_2   {\cal M}^{ini}_2}
 { \displaystyle   1 + e^{-t e_1} ( s_1 {\cal M}^{ini}_1 +s_2 {\cal M}^{ini}_2 ) 
   + e^{-t 2 e_1 }  s_1  s_2 { \cal C}^{ini}_{1,2}  }  
  \nonumber \\
\sum_{S_1=\pm 1} \sum_{S_2=\pm 1} S_2 B^{[0,t]}(S_1,S_2)    
&& =  \frac{1}{4}\sum_{s_1=\pm}  \sum_{s_2=\pm} 
\frac{ \displaystyle  {\cal M}^{ini}_2+  e^{-t e_1 }  (s_1 { \cal C}^{ini}_{1,2} +s_2 )  
+    e^{-t 2 e_1 }  s_1 s_2 {\cal M}^{ini}_1   }
 { \displaystyle   1 + e^{-t e_1} ( s_1 {\cal M}^{ini}_1 +s_2 {\cal M}^{ini}_2 ) 
   + e^{-t 2 e_1 }  s_1  s_2 { \cal C}^{ini}_{1,2}  }
\label{B0reconstructedbackwardspinsS1S2m1M2}
\end{eqnarray}
and the correlation between them
  \begin{eqnarray}
\sum_{S_1=\pm 1} \sum_{S_2=\pm 1} S_1 S_2 B^{[0,t]}(S_1,S_2) 
    = \frac{1}{4} \sum_{s_1=\pm}  \sum_{s_2=\pm} 
\frac{ \displaystyle 
\left[ { \cal C}^{ini}_{1,2}+  e^{-t e_1 }  (s_1 {\cal M}^{ini}_2 +s_2 {\cal M}^{ini}_1)  
+    e^{-t 2 e_1 }  s_1 s_2  \right] }
 { \displaystyle   1 + e^{-t e_1} ( s_1 {\cal M}^{ini}_1 +s_2 {\cal M}^{ini}_2 ) 
   + e^{-t 2 e_1 }  s_1  s_2 { \cal C}^{ini}_{1,2}  }
\label{B0reconstructedbackwardspinsS1S2C12}
\end{eqnarray}

The difference between $ B^{[0,t]}(S_1,S_2)$ of Eq. \ref{B0reconstructedbackwardspinsS1S2}
and its asymptotic limit $P^{ini}(S_1,S_2) $
is governed by the exponential decay of Eq. \ref{B0reconstructedbackwarddifferenceleadingspins}
  \begin{eqnarray}
B^{[0,t]}(S_1,S_2) - P^{ini}(S_1,S_2)  
&& = \frac{P^{ini}(S_1,S_2)}{4} \sum_{s_1=\pm}  \sum_{s_2=\pm} 
\frac{ \displaystyle    e^{-t e_1 }  [s_1 (S_1-{\cal M}^{ini}_1 ) +s_2 (S_2-{\cal M}^{ini}_2) ]  
+    e^{-t 2 e_1 }  s_1 s_2 (S_1  S_2 - { \cal C}^{ini}_{1,2} ) }
 { \displaystyle   1 + e^{-t e_1} ( s_1 {\cal M}^{ini}_1 +s_2 {\cal M}^{ini}_2 ) 
   + e^{-t 2 e_1 }  s_1  s_2 { \cal C}^{ini}_{1,2}  }
\nonumber \\
&&   \opsimeq_{t \to + \infty}
 e^{-t 2 e_1}
 P^{ini}(S_1,S_2)
  \left[{\cal M}^{ini}_1 ({\cal M}^{ini}_1 -S_1) +{\cal M}^{ini}_2({\cal M}^{ini}_2 -S_2) \right]  
\label{B0reconstructedbackwardspinsS1S2diff}
\end{eqnarray}
with the corresponding asymptotic behaviors for the two magnetizations
  \begin{eqnarray}
\sum_{S_1=\pm 1} \sum_{S_2=\pm 1} S_1 
 B^{[0,t]}(S_1,S_2) - M^{ini}_1
   \opsimeq_{t \to + \infty}
 e^{-t 2 e_1}
  \left[{\cal M}^{ini}_1 (({\cal M}^{ini}_1)^2 -1) +{\cal M}^{ini}_2({\cal M}^{ini}_1{\cal M}^{ini}_2 -{ \cal C}^{ini}_{1,2}) \right]  
  \nonumber \\
\sum_{S_1=\pm 1} \sum_{S_2=\pm 1} 
 S_2 B^{[0,t]}(S_1,S_2) - M^{ini}_2
   \opsimeq_{t \to + \infty}
 e^{-t 2 e_1}
  \left[{\cal M}^{ini}_1 ({\cal M}^{ini}_1{\cal M}^{ini}_2 -{ \cal C}^{ini}_{1,2}) +{\cal M}^{ini}_2(({\cal M}^{ini}_2)^2 -1) \right]  
\label{B0reconstructedbackwardspinsS1S2diffmagne}
\end{eqnarray}
and for the correlation
  \begin{eqnarray}
\sum_{S_1=\pm 1} \sum_{S_2=\pm 1} S_1 S_2 B^{[0,t]}(S_1,S_2) - { \cal C}^{ini}_{1,2}
   \opsimeq_{t \to + \infty}
 e^{-t 2 e_1}
  \left[{\cal M}^{ini}_1 ({\cal M}^{ini}_1 { \cal C}^{ini}_{1,2}-{\cal M}^{ini}_2) +{\cal M}^{ini}_2({\cal M}^{ini}_2{ \cal C}^{ini}_{1,2} -{\cal M}^{ini}_1) \right]  
\label{B0reconstructedbackwardspinsS1S2diffcorre}
\end{eqnarray}


\subsection{ Toy model when $P^{ini}(\vec S)$ corresponds to the one-dimensional Ising model with couplings $J_{n=1,..,N}$ }

As a simple illustrative example, let us consider the case where the initial distribution $P^{ini}(S_1,S_2,..,S_N) $
of the $N$ spins $S_{n=1,..,N}$ corresponds to the Boltzmann distribution 
of the one-dimensional Ising model involving 
the $N$ arbitrary couplings $J_{n=1,..,N}$ (in amplitudes and in signs) and the given boundary external spin $S_0=+1$
 \begin{eqnarray}
P^{ini}(S_1,S_2,..,S_N) && \equiv \frac{1}{Z_N}  e^{ \displaystyle   \sum_{n=1}^{N}  J_n S_{n-1} S_{n} }
\nonumber \\
&& = \frac{1}{Z_N}  e^{ \displaystyle J_1 S_1+   J_2 S_1 S_2+   J_3 S_2 S_3 + ...J_{N-1} S_{N-2} S_{N-1} +    J_N S_{N-1} S_N }
  \label{1dIsing}
\end{eqnarray}


\subsubsection{ Magnetizations ${\cal M}^{ini}_i  $ and correlations ${\cal C}^{ini}_{i_1,i_2,...,i_K}  $ of the initial distribution $P^{ini}(S_1,S_2,..,S_N) $}

The rewriting of Eq. \ref{1dIsing}
 as the product
 \begin{eqnarray}
&& P^{ini}(S_1,S_2,..,S_N)  
= \frac{1}{Z_N}     \prod_{n=1}^{N} \left[ \cosh J_n +   S_{n-1} S_{n} \sinh J_n  \right] 
= \frac{1}{Z_N}   \left(  \prod_{i=1}^{N}  \cosh J_i   \right)
 \prod_{n=1}^{N} \left[ 1 +   S_{n-1} S_{n} \tanh J_n  \right] 
   \label{1dIsingprod}
 \\ && 
= \frac{1}{Z_N}   \left(  \prod_{i=1}^{N}  \cosh J_i   \right)
 \left[ 1 +   S_1 \tanh J_1  \right]  \left[ 1 +   S_1 S_2 \tanh J_2  \right]  ... \left[ 1 +   S_{N-2} S_{N-1} \tanh J_{N-1}  \right] 
  \left[ 1 +   S_{N-1} S_N \tanh J_N  \right] 
\nonumber
\end{eqnarray}
is useful to evaluate the partition function $Z_N$ ensuring the normalization of $P^{ini}(S_1,S_2,..,S_N) $ 
 \begin{eqnarray}
Z_N && =  \left(  \prod_{i=1}^{N}  \cosh J_i   \right) \sum_{S_1=\pm 1} \sum_{S_2=\pm 1} ... \sum_{S_N=\pm 1} 
 \left[ 1 +   S_1 \tanh J_1  \right]  \left[ 1 +   S_1 S_2 \tanh J_2  \right]  ... 
  \left[ 1 +   S_{N-1} S_N \tanh J_N  \right] 
  \nonumber \\
  && = 2^N \left(  \prod_{i=1}^{N}  \cosh J_i   \right)
  \label{1dIsingpartition}
\end{eqnarray}
so that Eq. \ref{1dIsingprod} reduces to
  \begin{eqnarray}
P^{ini}(S_1,S_2,..,S_N) && = \frac{1}{2^N}  \prod_{n=1}^{N} \left[ 1 +   S_{n-1} S_{n} \tanh J_n  \right] 
\nonumber \\
&& = \frac{1}{2^N}  
 \left[ 1 +   S_1 \tanh J_1  \right]  \left[ 1 +   S_1 S_2 \tanh J_2  \right]  ... \left[ 1 +   S_{N-2} S_{N-1} \tanh J_{N-1}  \right] 
  \left[ 1 +   S_{N-1} S_N \tanh J_N  \right] 
\nonumber
  \label{1dIsingprodth}
\end{eqnarray}
 
 One can thus write the
 $N$ magnetizations ${\cal M}^{ini}_{i=1,..,N}$ of the $N$ spins $S_{i=1,..,N} $ 
   \begin{eqnarray}
{\cal M}^{ini}_i \equiv   \sum_{S_1=\pm 1} \sum_{S_2=\pm 1} ... \sum_{S_N=\pm 1}  S_i P^{ini}(S_1,S_2,..,S_N) 
  = \prod_{n=1}^i \tanh J_n
  \label{1dIsingmagnetizationsmi}
\end{eqnarray}
and the $\frac{N(N-1)}{2}$ two-spin-correlations ${\cal C}^{ini}_{i_1,i_2}$ with $1 \leq i_1<i_2 \leq N$
    \begin{eqnarray}
{\cal C}^{ini}_{i_1,i_2} \equiv   \sum_{S_1=\pm 1} \sum_{S_2=\pm 1} ... \sum_{S_N=\pm 1}  S_{i_1} S_{i_2} P^{ini}(S_1,S_2,..,S_N) 
  = \prod_{n=i_1+1}^{i_2} \tanh J_n
  \label{1dIsingCorreci1i2}
\end{eqnarray}
that are actually sufficient to reconstruct all the higher correlations as follows.
The three-spin-correlations ${\cal C}^{ini}_{i_1,i_2,i_3}$ with $1 \leq i_1<i_2<i_3 \leq N$ read
    \begin{eqnarray}
{\cal C}^{ini}_{i_1,i_2,i_3} && \equiv   \sum_{S_1=\pm 1} \sum_{S_2=\pm 1} ... \sum_{S_N=\pm 1}  S_{i_1} S_{i_2} S_{i_3}P^{ini}(S_1,S_2,..,S_N) 
 \nonumber \\ &&   = \left( \prod_{n_1=1}^{i_1} \tanh J_{n_1} \right) \left( \prod_{n_2=i_2+1}^{i_3} \tanh J_{n_2} \right)
\equiv {\cal M}^{ini}_{i_1}{\cal C}^{ini}_{i_2,i_3}
  \label{1dIsingCorreci1i2i3}
\end{eqnarray}
while the four-spin-correlations ${\cal C}^{ini}_{i_1,i_2,i_3,i_K}$ with $1 \leq i_1<i_2<i_3<i_4 \leq N$ reduce to
    \begin{eqnarray}
{\cal C}^{ini}_{i_1,i_2,i_3,i_4} 
&& \equiv   \sum_{S_1=\pm 1} \sum_{S_2=\pm 1} ... \sum_{S_N=\pm 1}  S_{i_1} S_{i_2} S_{i_3}S_{i_4}P^{ini}(S_1,S_2,..,S_N) 
\nonumber \\
&&  = \left(\prod_{n_1=i_1+1}^{i_2} \tanh J_n  \right) \left( \prod_{n=i_1+1}^{i_2} \tanh J_{n_2} \right)
  \equiv {\cal C}^{ini}_{i_1,i_2} {\cal C}^{ini}_{i_3,i_4}
  \label{1dIsingCorreci1i2i3i4}
\end{eqnarray}
 More generally, the correlations ${\cal C}^{ini}_{i_1,i_2,...,i_{2K}}$ between $(2K)$ spins with $1 \leq i_1<i_2<...<i_{2K} \leq N$ reduce to the product of the $K$ two-spin-correlations ${\cal C}^{ini}_{i_{2k-1},i_{2k}} $
     \begin{eqnarray}
{\cal C}^{ini}_{i_1,i_2,...,i_{2K}} = \prod_{k=1}^K {\cal C}^{ini}_{i_{2k-1},i_{2k}} 
  \label{1dIsingCorreci2Keven}
\end{eqnarray}
while the correlations ${\cal C}^{ini}_{i_1,i_2,...,i_{2K+1}}$ between $(2K+1)$ spins with $1 \leq i_1<i_2<...<i_{2K+1} \leq N$ reduce to
     \begin{eqnarray}
{\cal C}^{ini}_{i_1,i_2,...,i_{2K+1}} ={\cal M}^{ini}_{i_1} \prod_{k=1}^{K} {\cal C}^{ini}_{i_{2k},i_{2k+1}} 
  \label{1dIsingCorreci2Km1odd}
\end{eqnarray}


\subsubsection{ Convergence of the forward solution $P(\vec s,t) $ }

The magnetizations ${\cal M}^{ini}_i  $ and the correlations ${\cal C}^{ini}_{i_1,i_2,...,i_K}  $ computed in the previous subsection
can be plugged into Eq. \ref{Probafromcorredynexpli} to obtain the 
spectral decomposition of the forward solution 
 \begin{eqnarray}
P(\vec s,t) && 
  =   \frac{1}{2^N} \left( 1 +  e^{- t e_1 } \sum_{i=1}^N s_i {\cal M}^{ini}_i
  +  e^{- t 2 e_1 } \sum_{1 \leq i_1 < i_2 \leq N} s_{i_1} s_{i_2}{\cal C}^{ini}_{i_1,i_2}
  + \sum_{K=3}^N e^{- t K e_1 } \sum_{1 \leq i_1 < i_2 ..<i_K \leq N} s_{i_1} s_{i_2}..s_{i_K}{\cal C}^{ini}_{i_1,i_2,..,i_K}
   \right)
   \nonumber \\
   &&  =   \frac{1}{2^N} \left( 1 +  e^{- t e_1 } \sum_{i=1}^N s_i \left(  \prod_{n=1}^i \tanh J_n \right)
  +  e^{- t 2 e_1 } \sum_{1 \leq i_1 < i_2 \leq N} s_{i_1} s_{i_2} \left(\prod_{n=i_1+1}^{i_2} \tanh J_n \right)
  + ...\right)
\label{Probafromcorredynexplispins}
\end{eqnarray}


\subsubsection{ Convergence properties of the  reconstructive backward dynamics  }

The asymptotic behavior of Eq. \ref{B0reconstructedbackwarddifferenceleadingspins}
reads
  \begin{eqnarray}
 B^{[0,t]}(\vec S) 
 && \opsimeq_{t \to + \infty}   
  P^{ini}(\vec S)   
   \left[ 1+e^{- t 2 e_1 }
\sum_{i=1}^N  {\cal M}_i^{ini}  \left[  {\cal M}_i^{ini} -  S_i \right]
    +...\right]
 \nonumber \\
&&     \opsimeq_{t \to + \infty}   
  P^{ini}(\vec S)  
   \left[ 1+e^{- t 2 e_1 }
\sum_{i=1}^N \left(  \prod_{n_1=1}^i \tanh J_{n_1} \right)  \left[ \left(  \prod_{n_2=1}^i \tanh J_{n_2} \right) -  S_i \right]
    +...\right]
\label{B0reconstructedbackwarddifferenceleadingspins1d}
\end{eqnarray}

The convergence of the magnetization of the single pixel $j$
of Eq. \ref{B0reconstructedbackwarddifferenceleadingobsleftl1spins} 
requires the valuation of the connected correlation between two spins $S_{i_1}$ and $S_{i_2}$
for $1 \leq i_1<i_2 \leq N$ using Eqs \ref{1dIsingmagnetizationsmi}
and \ref{1dIsingCorreci1i2}
 \begin{eqnarray}
  { \cal C}^{ini}_{i_1,i_2} - {\cal M}^{ini}_{i_1}  {\cal M}^{ini}_{i_2}  && =  \prod_{n=i_1+1}^{i_2} \tanh J_n
  - \left( \prod_{n_1=1}^{i_1} \tanh J_{n_1} \right)  \left(  \prod_{n_2=1}^{i_2} \tanh J_{n_2} \right)
  \nonumber \\
  && = \left[ 1-  \left( \prod_{n_1=1}^{i_1} \tanh J_{n_1} \right)^2\right]  \prod_{n=i_1+1}^{i_2} \tanh J_n
  =  \left[ 1-  \left( {\cal M}^{ini}_{i_1} \right)^2\right] { \cal C}^{ini}_{i_1,i_2}
\label{cijconnectedspins1d}
\end{eqnarray}
so that Eq. \ref{B0reconstructedbackwarddifferenceleadingobsleftl1spins}
becomes
\begin{small}
  \begin{eqnarray}
&& \sum_{\vec S}  S_j B^{[0,t]}(\vec S) -  {\cal M}^{ini}_j 
 \opsimeq_{t \to + \infty}   
-   e^{- t 2 e_1 }  \left(  {\cal M}^{ini}_j 
\left[  1 - ({\cal M}^{ini}_j )^2  
   \right]  
+ \sum_{i =1 }^{j-1}  {\cal M}^{ini}_i 
\left[   { \cal C}^{ini}_{i,j} - {\cal M}^{ini}_i  {\cal M}^{ini}_j     \right]  
+ \sum_{i = j +1}^N  {\cal M}^{ini}_i 
\left[   { \cal C}^{ini}_{j,i} - {\cal M}^{ini}_j  {\cal M}^{ini}_i     \right]  
   \right)
    +...  
     \nonumber \\
 &&
  \opsimeq_{t \to + \infty}      -   e^{- t 2 e_1 }  \left(  {\cal M}^{ini}_j \left[  1 - ({\cal M}^{ini}_j )^2     \right]  
+ \sum_{i =1 }^{j-1}   
\left[ 1-  \left( {\cal M}^{ini}_{i} \right)^2\right] {\cal M}^{ini}_i { \cal C}^{ini}_{i,j}
+  \left[ 1-  \left( {\cal M}^{ini}_{j} \right)^2\right] \sum_{i = j +1}^N  {\cal M}^{ini}_i  { \cal C}^{ini}_{j,i}   
   \right)
    +...  
       \nonumber \\    
    &&
  \opsimeq_{t \to + \infty}      -   e^{- t 2 e_1 }  \left(  {\cal M}^{ini}_j \left[  1 - ({\cal M}^{ini}_j )^2     \right]  
+ \sum_{i =1 }^{j-1}   
\left[ 1-  \left( {\cal M}^{ini}_{i} \right)^2\right] {\cal M}^{ini}_j 
+  \left[ 1-  \left( {\cal M}^{ini}_{j} \right)^2\right] \sum_{i = j +1}^N  {\cal M}^{ini}_j  \left( { \cal C}^{ini}_{j,i}   \right)^2
   \right)
    +...      
           \nonumber \\    
    &&
  \opsimeq_{t \to + \infty}      -   e^{- t 2 e_1 }  {\cal M}^{ini}_j
  \left(   \left[  1 - ({\cal M}^{ini}_j )^2     \right]  
+ \sum_{i =1 }^{j-1}   \left[ 1-  \left( {\cal M}^{ini}_{i} \right)^2\right] 
+  \left[ 1-  \left( {\cal M}^{ini}_{j} \right)^2\right] \sum_{i = j +1}^N    \left( { \cal C}^{ini}_{j,i}   \right)^2
   \right)
    +...      
\label{B0reconstructedbackwarddifferenceleadingobsleftl1spins1d}
\end{eqnarray}
\end{small}
For instance the special case $j=1$ reduces to
 \begin{eqnarray}
&& \sum_{\vec S}  S_1 B^{[0,t]}(\vec S) -  {\cal M}^{ini}_1 
 \opsimeq_{t \to + \infty}   
  -   e^{- t 2 e_1 }  {\cal M}^{ini}_1
  \left(   \left[  1 - ({\cal M}^{ini}_1 )^2     \right]  
+ 0
+  \left[ 1-  \left( {\cal M}^{ini}_1 \right)^2\right] \sum_{i = 2}^N    \left( { \cal C}^{ini}_{1,i}   \right)^2   \right)
    +...      
     \nonumber \\
 &&
  \opsimeq_{t \to + \infty}     
    -   e^{- t 2 e_1 } \tanh J_1 \left[  1 - \tanh^2 J_1     \right]
  \left(   1+ \sum_{i = 2}^N    \left( \prod_{n=2}^i \tanh J_n    \right)^2   \right)
    +...   
  \label{B0reconstructedbackwarddifferenceleadingobsleftl1spins1dj1}
\end{eqnarray}

Finally, the convergence for the correlation between the $K=2$ spins $S_1$ and $S_2$ of Eq. \ref{B0reconstructedbackwarddifferenceleadingobsleftspins12}
can be evaluated using using Eqs \ref{1dIsingmagnetizationsmi},
 \ref{1dIsingCorreci1i2} and \ref{1dIsingCorreci1i2i3}
 \begin{small}
  \begin{eqnarray}
&& \sum_{\vec S}  S_1 S_2  B^{[0,t]}(\vec S) -  { \cal C}^{ini}_{1,2}
 \opsimeq_{t \to + \infty}   
-   e^{- t 2 e_1 }  \left(  {\cal M}^{ini}_1 
\left[  {\cal M}^{ini}_2 - {\cal M}^{ini}_1 { \cal C}^{ini}_{1,2}  
   \right]  
    + {\cal M}^{ini}_2 
\left[  {\cal M}^{ini}_1 - {\cal M}^{ini}_2 { \cal C}^{ini}_{1,2}  
   \right] 
   + \sum_{i =3 }^N  {\cal M}^{ini}_i 
\left[   { \cal C}^{ini}_{1,2,i} - { \cal C}^{ini}_{1,2} {\cal M}^{ini}_i 
   \right]  \right)     +...    
    \nonumber \\
 &&   \opsimeq_{t \to + \infty}   
-   e^{- t 2 e_1 }  \left(  0    + \tanh^2 J_1 \tanh J_2 
\left[  1 -  \tanh^2 J_2      \right] 
   +\left[   \frac{1}{\tanh J_2}- \tanh J_2   \right] \sum_{i =3 }^N  \left( \prod_{n=1}^i \tanh J_n \right)^2
  \right)     +...      
\label{B0reconstructedbackwarddifferenceleadingobsleftspins121d}
\end{eqnarray}
\end{small}


\section{ Application to generative diffusion models  }

\label{sec_diff}

In this section, we focus on the case where each pixel $n=1,..,N$ can
be in an infinite number of configurations described by the continuous variable $x_n \in ]x_-,x_+[$,
while the forward dynamics is governed by a Fokker-Planck differential generator.

\subsection{ Spectral properties of the forward Pearson diffusion for a single pixel }

\subsubsection{ Reminder on diffusion processes on the interval $]x_-,x_+[ $ }

The Fokker-Planck dynamics 
for the probability density $p_t(x)$ to be at $x$ at time $t$
corresponds to
the continuity equation
\begin{eqnarray}
 \partial_t p_t(x)    && =   -  \partial_{x}   j_t(x)
 \nonumber \\
  j_t(x) && \equiv F(x)   p_t(x) - D (x)  \partial_{x} p_t(x)
\label{fokkerplanck}
\end{eqnarray}
where the current $j_t(x) $ associated to $p_t(x) $
involves the Fokker-Planck force $F(x) $ and the diffusion coefficient $D(x)$,
while it should vanish at the two boundaries $x_-$ and $x_+$
\begin{eqnarray}
j_t(x_-)  =0 =j_t(x_+) 
\label{jboundaries}
\end{eqnarray}
in order to conserve
 the total probability $\int_{x_-}^{x_+} dx p_t(x)=1 $.
 The steady state $p_*(x)$ associated to the vanishing steady current $ j_*(x)=0 $
 for any $x$ (this represents the continuous version
  of the discrete-configuration detailed-balance condition of Eq. \ref{DBdef})
\begin{eqnarray}
0=  j_*(x) = F(x)   p_*(x) - D (x)  \partial_{x} p_*(x)
\label{jsteadyzero}
\end{eqnarray}
reads in terms of an appropriate reference position $x_0 \in ]x_-,x_+[$ 
\begin{eqnarray}
  p_*(x)  = \frac{ e^{ \int_{x_0}^x dy \frac{F(y)}{D(y)}} }
  {  \int_{x_-}^{x_+} dz e^{\int_{x_0}^z dy' \frac{F(y')}{D(y')}  }}
 \label{steadyeq}
\end{eqnarray}
The similarity transformation of Eq. \ref{similarity} transforms the differential
generator $w$ of the Fokker-Planck dynamics of Eq. \ref{fokkerplanck}
\begin{eqnarray}
 w    =  - \partial_{x}  \left[ F(x)  - D(x) \partial_x \right] 
\label{fokkerplanckgene}
\end{eqnarray}
into the quantum symmetric Hamiltonian 
\begin{eqnarray}
H  =-  \frac{1}{  \sqrt{p_* (x) }} w  \sqrt{p_* (x) } 
=   - \frac{ \partial  }{\partial x} D(x) \frac{ \partial  }{\partial x} +  \frac{ F^2(x) }{4 D(x) } + \frac{F'(x)}{2}
\label{similarityPearson}
\end{eqnarray}
In order to be more concrete and to have a simple physical interpretation for the left eigenvectors
that play a major role in the general framework described in sections \ref{sec_forward} and \ref{sec_backward},
we will now focus on the special case of Pearson diffusions. 


\subsubsection{ Reminder on the specific properties of Pearson diffusions with finite steady moments }

Among the one-dimensional diffusions involving an arbitrary force $F(x)$ and an arbitrary diffusion coefficient $D(x)$,
 the Pearson family where the force $F(x)$ is at most linear,
 where the diffusion coefficient $D(x) $ is at most quadratic,
and where the interval $]x_-,x_+[$ corresponds to the maximal interval 
where the diffusion coefficient remains positive $D(x) > 0$
is characterized by many simplifications
\cite{pearson_wong,pearson_class,pearson2012,PearsonHeavyTailed,pearson2018,c_pearson}.
In particular, the dynamical equation for the moment of order $k$ 
\begin{eqnarray}
m_k(t)\equiv \int_{x_-}^{x_+} dx x^k p_t(x)
\label{mkt}
\end{eqnarray}
only involves the moment $m_k(t)$ itself and at most the two lower moments $m_{k-1}(t)$ and $m_{k-2}(t)$.
In the present paper, we will focus on 
the Pearson 'half-family' where the moments $m_k^*$ of the steady state $p_*(x)$ of Eq. \ref{steadyeq}
are finite for any $k=1,2,..+\infty$ 
\begin{eqnarray}
m_k^* \equiv \int_{x_-}^{x_+} dx x^k p_*(x) < + \infty \ \ \ \text{ for } \ \ k=1,2,..+\infty
\label{mkstar}
\end{eqnarray}
Then the left eigenvectors $l_{\alpha}(x)$ of the 
differential generator  $w$ of Eq. \ref{fokkerplanckgene}
that correspond to the right eigenvectors of the adjoint differential operator $w^{\dagger}$
\begin{eqnarray}
 - e_{\alpha} l_{\alpha}(x)   =  w^{\dagger}  l_{\alpha}(x) 
 =  F(x) \partial_{x} l_{\alpha}(x) +\partial_x \bigg( D(x)  \partial_{x} l_{\alpha}(x)  \bigg)
\label{leftwdagger}
\end{eqnarray}
and that form the orthonormal family of Eq. \ref{ortholeft}
with respect to the steady state measure $p_*(x)$
  \begin{eqnarray}
  \delta_{\alpha,\beta}   = \int_{x_-}^{x_+} dx    l_{\alpha} (x)  l_{\beta}(x) p_*(x)
\label{ortholeftpearson}
\end{eqnarray}
reduce to polynomials of degree $\alpha=0,1,..+\infty$ starting with $l_0(x)=1$ 
\begin{eqnarray}
 l_{\alpha}(x) = \sum_{k=0}^{\alpha} \Upsilon_{\alpha, k} x^k 
 = \Upsilon_{\alpha, \alpha } x^{\alpha} + \Upsilon_{\alpha, \alpha-1 } x^{\alpha-1}
 + ... +\Upsilon_{\alpha,1 } x + \Upsilon_{\alpha,0 }
 \label{leftpolynomial}
\end{eqnarray}
The associated observable $l_{\alpha}^{av}(t) $ of Eq. \ref{leftobservable}
that displays the simple exponential dynamics in $e^{-t e_{\alpha} } $
corresponds to the linear combination of moments $m_k(t)$ of Eq. \ref{mkt} with $k=0,..,\alpha$
\begin{eqnarray}
l_{\alpha}^{av}(t) && \equiv \int_{x_-}^{x_+} dx  l_{\alpha}(x)  p_t(x) = e^{-t e_{\alpha} }    l^{av}_{\alpha}(0 ) 
\nonumber \\
&& = \int_{x_-}^{x_+} dx \left[  \sum_{k=0}^{\alpha} \Upsilon_{\alpha k} x^k  \right]  p_t(x)
=  \sum_{k=0}^{\alpha} \Upsilon_{\alpha k} m_k(t)
 \label{lnavpearson}
\end{eqnarray}
Reciprocally, the convergence of the moment $ m_k(t)$ towards its steady value $m_k^* $
involves the $k$ exponentials $e^{-t e_{\alpha} } $ with $\alpha=1,2.,.,k$.

The orthonormalisation relations of Eq. \ref{ortholeftpearson} yield that the 
coefficients $\Upsilon_{\alpha k} $ of Eq. \ref{leftpolynomial}
can be rewritten in terms of the steady moments $m_k^* $ of Eq. \ref{mkstar}.
For instance the first eigenvector $l_1(x) $ reads in terms of the two first steady moments $m^*_{k=1,2} $ 
\begin{eqnarray}
l_1(x) = \Upsilon_{1,1 } x + \Upsilon_{1,0 }
= \frac{ x - m_1^*}{\sqrt{ m_2^* - \left(m_1^*\right)^2 } }
 \label{l1pearson}
\end{eqnarray}
so that the corresponding observable $l_1^{av}(t) $ of Eq. \ref{lnavpearson}
has a very simple interpretation in terms of the first moment $m_1(t)$
\begin{eqnarray}
l_1^{av}(t) && \equiv \int_{x_-}^{x_+} dx  l_1(x)  p_t(x) = e^{-t e_1 }    l^{av}_{\alpha}(0 ) 
\nonumber \\
&& = \int_{x_-}^{x_+} dx \left[   \frac{ x - m_1^*}{\sqrt{ m_2^* - \left(m_1^*\right)^2 } }  \right]  p_t(x)
=   \frac{ m_1(t) - m_1^*}{\sqrt{ m_2^* - \left(m_1^*\right)^2 } }
 \label{l1avpearson}
\end{eqnarray}

The spectral decomposition of Eq. \ref{spectralpixel}
for the propagator involves the infinite series of the polynomials $l_{ \alpha} $
  \begin{eqnarray}
 \langle \vec x \vert e^{  t w} \vert \vec X \rangle
&& = p_*( \vec x) \bigg( 1+ \sum_{\alpha=1}^{+\infty} e^{- t e_{\alpha} } l_{ \alpha}(\vec x) l_{ \alpha}(\vec X) \bigg)
\label{fpropagatorspectralonlyleftpearson}
\end{eqnarray}
while the energy spectrum $e_{\alpha} $ can be either linear or quadratic in $\alpha$
as recalled for the three basic examples on the various intervals 
$]-\infty,+\infty[ $, $]0,+\infty[ $ and $]0,1[ $ described in the three following subsections.


\subsubsection{ Example of the Ornstein-Uhlenbeck process on $x \in ]-\infty,+\infty[$ with the Gaussian distribution as steady state  }

The Pearson diffusion on $x \in ]-\infty,+\infty[$ with the constant diffusion coefficient 
\begin{eqnarray}
 D(x)   = \frac{1}{2} 
\label{OUD}
\end{eqnarray}
and the linear restoring force of parameter $\omega$ towards the origin
\begin{eqnarray}
 F(x)  = - \omega x
\label{OUforce}
\end{eqnarray}
corresponds to the famous Ornstein-Uhlenbeck process that is
 the most standard example used in the field of Generative diffusion models.
 Its explicit Gaussian propagator  
  \begin{eqnarray}
\langle x \vert e^{  t w} \vert X \rangle 
 = \sqrt {\frac{\omega}{\pi (1-e^{-2 \omega t} ) }} 
 e^{ \displaystyle - \omega \frac{ \left( x - X e^{-\omega t} \right)^2}{(1-e^{-2 \omega t} )} }
\label{OUpropagator}
\end{eqnarray}
converges for $t \to + \infty$ towards the Gaussian steady state
 \begin{eqnarray}
p_*(x)  
 = \sqrt {\frac{\omega}{\pi  }} e^{  - \omega x^2 }
\label{OUsteady}
\end{eqnarray}
The similarity transformation of Eq. \ref{similarityPearson}
leads to the well-known quantum Hamiltonian of an harmonic oscillator 
\begin{eqnarray}
H =-  \frac{1}{  \sqrt{p_* (x) }} w  \sqrt{p_* (x) } 
=  - \frac{1}{2} \partial_x^2 + \frac{\omega^2}{2} x^2 - \frac{\omega}{2}
\label{similarityOU}
\end{eqnarray}
with its linear spectrum
\begin{eqnarray}
e_{\alpha}= \omega \alpha \ \ \ \text{ with } \ \ \alpha=0,1,2,...+\infty
\label{spectreOU}
\end{eqnarray}
while the left eigenvectors $l_{\alpha}(x)$ involve the Hermite polynomials.


\subsubsection{ Example of the Square-Root process on $x \in ]0,+\infty[$ with the $Gamma$-distribution as steady state  }

The Pearson diffusion on $x \in ]0,+\infty[$ with the linear diffusion coefficient 
\begin{eqnarray}
 D(x)   =x   
\label{gammaD}
\end{eqnarray}
and the linear force parametrized by the two coefficients $\mu>0$ and $\lambda>0$  
\begin{eqnarray}
 F(x) && = (\mu-1) - \lambda x 
\label{gammaF}
\end{eqnarray}
is
called either the Square-Root process \cite{dufresne} or the Cox-Ingersoll-Ross process \cite{CIR}.
The steady state corresponds to the Gamma-distribution 
\begin{eqnarray}
p_*(x) = \frac{\lambda^{\mu}}{ \Gamma(\mu)} x^{\mu-1}e^{- \lambda x} \ \ \ \ \ {\rm for } \ \ x \in ]0,+\infty[
\label{gamma}
\end{eqnarray}
with the simple moments
\begin{eqnarray}
m_k^* =    \int_0^{+\infty} dx x^k p_*(x) 
 = \frac{\Gamma(\mu+k)}{ \lambda^k \Gamma(\mu)} 
 \label{mkgammasteady}
\end{eqnarray}
The left eigenvectors $l_{\alpha}$ involve the Laguerre polynomials 
\cite{pearson_wong,pearson_class,pearson2012,PearsonHeavyTailed,pearson2018},
while the eigenvalues $e_{\alpha} $ are linear in $\alpha$ 
\begin{eqnarray}
 e_{\alpha}  = \lambda \alpha \ \ \ \text{ for } \ \ \alpha=0,1,2,..+\infty
  \label{ekmkgamma}
\end{eqnarray}


\subsubsection{ Example of the Jacobi process on $x \in ]0,1[$ with the $Beta$-distribution as steady state  }

The Pearson diffusion on the finite interval $x \in ]0,1[$ with the quadratic diffusion coefficient 
\begin{eqnarray}
 D(x)   =x (1-x)  
\label{betaD}
\end{eqnarray}
and the linear force parametrized by the two coefficients $a>0$ and $b>0$  
\begin{eqnarray}
 F(x) && =  (a-1) - (a+b-2) x
\label{beta}
\end{eqnarray}
is called the Jacobi process.
The steady state corresponds to the Beta-distribution 
\begin{eqnarray}
p_*(x) =
\frac{\Gamma(a+b)}{ \Gamma(a) \Gamma(b) } x^{a-1} (1-x)^{b-1}
 \ \ \ {\rm for } \ x \in ]0,1[
\label{jacobi}
\end{eqnarray}
with the moments
\begin{eqnarray}
m_k^*  =  \int_0^{1} dx x^{k} p_*(x) 
  = \frac{\Gamma(a+k) \Gamma(a+b)}{ \Gamma(a) \Gamma(a+b+k) }
   \label{mksteadyjacobi}
\end{eqnarray}
The left eigenvectors $l_{\alpha}$ 
involve the Jacobi polynomials
(see  \cite{pearson_wong,pearson_class,pearson2012,PearsonHeavyTailed,pearson2018}
and references therein)
while the eigenvalues $e_{\alpha} $ are quadratic in $\alpha$
\begin{eqnarray}
 \epsilon_{\alpha} = \alpha^2+ \alpha (  a+b -1)
  \label{ekmkjacobi}
\end{eqnarray}


\subsection{ Properties of forward dynamics for the $N$ pixels }

For the configuration $\vec x=(x_1,..,x_N)$ of the $N$ pixels, 
the diffusion process becomes $N$-dimensional,
with the spectral decomposition of Eq. \ref{factorizedpropagatorspectral}
 for the global propagator 
  \begin{eqnarray}
 \langle \vec x \vert e^{  t W} \vert \vec X \rangle
&& = P_*( \vec x) \bigg( 1+ \sum_{\vec \alpha \ne \vec 0} e^{- t E_{\vec \alpha} } L_{\vec \alpha}(\vec x) L_{\vec \alpha}(\vec X) \bigg)
\nonumber \\
&& = P_*( \vec x) \bigg( 1+ \sum_{\vec \alpha \ne \vec 0} 
\prod_{n=1}^N \left[ e^{- t e_{\alpha_n} } l_{\alpha_n}(x_n) l_{\alpha_n}(X_n) \right] \bigg)
\label{factorizedpropagatorspectralonlyleftOU}
\end{eqnarray}

In particular, the leading contribution of order $e^{-t e_1} $ in Eq. \ref{factorizedpropagatorspectralonlyleftfirstexcited}
governing
the convergence of Eq. \ref{factorizedpropagatorspectralonlyleftOU} towards the steady state
involve the first polynomial $l_1(x) $ of Eq. \ref{l1pearson}
  \begin{eqnarray}
 \langle \vec x \vert e^{  t W} \vert \vec X \rangle
&& = P_*( \vec x) \bigg[  1+  e^{- t e_1 } \sum_{n=1}^N l_1(x_n) l_1(X_n) +...\bigg]
\nonumber \\
&& = P_*( \vec x) \bigg[  1+  e^{- t e_1 } \sum_{n=1}^N  \frac{ (x_n - m_1^*)(X_n - m_1^* )}{ m_2^* - \left(m_1^*\right)^2  }  +...\bigg]
\label{factorizedpropagatorspectralonlyleftfirstexcitedN}
\end{eqnarray}

Here the observables $ L_{\vec \alpha}^{av}(t )   $ of Eq. \ref{leftobservableGlobal}
whose dynamics involve a single global energy $E_{\vec \alpha} $ 
  \begin{eqnarray}
 L_{\vec \alpha}^{av}(t )   \equiv \int d^N \vec x  L_{\alpha}(\vec x) P( \vec x, t)
= \left[ \prod_{n=1}^N\int_{x_-}^{x_+} dx_n  l_{\alpha_n}(x_n) \right] P( \vec x, t)
 =  L^{ini}_{\alpha} e^{- t E_{\vec \alpha} }
\label{leftobservableGlobaldiff}
\end{eqnarray}
starting from their initial values of Eq. \ref{Leftini}
 \begin{eqnarray}
 L^{ini}_{\vec \alpha} \equiv \int d^N \vec x  L_{\alpha}(\vec x) P^{ini}( \vec x)
= \left[ \prod_{n=1}^N\int_{x_-}^{x_+} dx_n  l_{\alpha_n}(x_n) \right] P^{ini}( \vec x) 
\label{Leftinidiff}
\end{eqnarray}
involve products over the $N$ pixels of the polynomials $l_{\alpha_n}(x_n)$ of degrees $\alpha_n\in \{0,1,..+\infty\}$ 
of Eq. \ref{leftpolynomial}.


\subsection{ Convergence properties of the reconstructive backward dynamics }

The reconstructed distribution of Eq. \ref{B0reconstructedbackward}
reads 
  \begin{eqnarray}
B^{[0,t]}(\vec X) &&  = P^{ini}(\vec X) 
\int d^N \vec x  P_*(\vec x) 
 \frac{ \displaystyle   1 + \sum_{\vec \alpha \ne \vec 0} e^{- t E_{\vec \alpha} } 
L_{\vec \alpha}(\vec x)  L_{\vec \alpha}(\vec X)  }
 { \displaystyle 1 + \sum_{\vec \beta \ne \vec 0} e^{- t E_{\vec \beta} } L_{\vec \beta}(\vec x) 
L_{\vec \beta}^{ini}   } 
\label{B0reconstructedbackwardspectralOU}
\end{eqnarray}
while the difference with the asymptotic limit $P^{ini}(\vec X) $
is governed by the leading exponential decay of Eq. \ref{B0reconstructedbackwarddifferenceleading}
  \begin{eqnarray}
B^{[0,t]}(\vec X) - P^{ini}(\vec X)
&&  
= P^{ini}(\vec X) 
\int d^N \vec x  P_*(\vec x) 
 \frac{ \displaystyle    \sum_{\vec \alpha \ne \vec 0} e^{- t E_{\vec \alpha} } 
L_{\vec \alpha}(\vec x)  \left[ L_{\vec \alpha}(\vec X) 
-    L_{\vec \alpha}^{ini}  \right] }
 { \displaystyle 1 + \sum_{\vec \beta \ne \vec 0} e^{- t E_{\vec \beta} } L_{\vec \beta}(\vec x) 
 L_{\vec \beta}^{ini}  } 
\nonumber \\
 &&   \opsimeq_{t \to + \infty}   
   e^{- t 2 e_1 } 
    P^{ini}(\vec X)   
\sum_{n=1}^N  \langle l_1^{[n]} \vert  P^{ini} \rangle \left[  \langle l_1^{[n]} \vert  P^{ini} \rangle -  l_1(X_n) \right]
    +...
\label{B0reconstructedbackwardspectraldifferenceleadingOU}
\end{eqnarray}
where on can use the explicit form of Eq. \ref{l1pearson}
for the first polynomial $l_1(.) $ to replace 
  \begin{eqnarray}
  \langle l_1^{[n]} \vert  P^{ini} \rangle  =  \int d^N \vec Y l_1(Y_n)  P^{ini} (\vec Y) 
  =
  \frac{ \int d^N \vec Y Y_n P^{ini} (\vec Y)- m_1^*}{\sqrt{m_2^* - \left(m_1^*\right)^2 } } 
\label{L1nPiniY}
\end{eqnarray}
and to obtain the final explicit result
 \begin{eqnarray}
B^{[0,t]}(\vec X) - P^{ini}(\vec X)
   \opsimeq_{t \to + \infty}   
   e^{- t 2 e_1 } 
    P^{ini}(\vec X)   
\sum_{n=1}^N   \frac{\left[  \int d^N \vec Y Y_n P^{ini} (\vec Y)- m_1^* \right]
 \left[   \int d^N \vec Z Z_n P^{ini} (\vec Z)- X_n \right] }{m_2^* - \left(m_1^*\right)^2  }   
    +...
\label{B0reconstructedbackwardspectraldifferenceleadingOUfinal}
\end{eqnarray}
from which one can derive the leading convergence of any observable ${\cal O}(\vec X)$ 
of the configuration $\vec X$ of the $N$ pixels.

\subsection{ Example: Ornstein-Uhlenbeck dynamics when $P^{ini}(\vec X)$ is a mixture of Gaussian distributions }

As a simple toy model to obtain analytical properties, many recent works have considered 
the Ornstein-Uhlenbeck dynamics with its Gaussian propagator of Eq. \ref{OUpropagator}
when the initial condition $P^{ini}(\vec X)$ is a mixture of delta functions or of Gaussian distributions
(see \cite{Biroli_largedim,Ambrogioni1,Ambrogioni2,Ambrogioni3,Biroli_dynamical,emergence,pathintegral,GeneNonEq,wasserstein,Ambrogioni4,Ambrogioni5,cumulants} and references therein),
in particular to analyze the spontaneous symmetry breaking that will take place between the many-valleys-initial-condition and the asymptotic single-valley-steady-state,
and to characterize the memorization with respect to the initial data.

In this subsection, it is thus interesting to apply the framework described 
in the previous sections to the case where the initial distribution   
 \begin{eqnarray}
P^{ini}( \vec X) = \sum_{a=1}^A \pi_a G^{[\vec \mu^{[a]}; {\bold \Sigma}^{[a]}]} (\vec X)
\label{mixture}
\end{eqnarray}
is a mixture with the normalized weights $\pi_a \in ]0,1[$ 
 \begin{eqnarray}
 \sum_{a=1}^A \pi_a =1
\label{pia}
\end{eqnarray}
of the multivariate Gaussian distributions $G^{[\vec \mu^{[a]}; {\bold \Sigma}^{[a]}]} (\vec X) $
of averaged values $\vec \mu^{[a]} $ and of $N \times N$ covariance matrices ${\bold \Sigma}^{[a] }$,
with the notation
\begin{eqnarray}
G^{[\vec \mu; {\bold \Sigma}]} (\vec X)
 \equiv \frac{1}{  \sqrt{(2\pi)^N \det({\bold \Sigma} ) } } 
 e^{ \displaystyle - \frac{1}{2} 
 \left( \langle \vec X \vert - \langle \vec \mu\vert\right)
  {\bold \Sigma}^{-1}
\left( \vert \vec X \rangle- \vert \vec \mu\rangle \right)
 }
\label{gaussmulti}
\end{eqnarray}


\subsubsection{ Averaged values and correlations when $P^{ini}( X_1,X_2,..,X_N)$ is a mixture of multivariate Gaussian distributions }

The averaged value $\mu_j^{ini}$ of $X_j$ in $P^{ini}( \vec X) $ of Eq. \ref{mixture} reduces to the weighted sum
of the $A$ averaged values $\vec \mu^{[a]} $
 \begin{eqnarray}
\mu_j^{ini} \equiv \int d^N \vec X \ X_j P^{ini}( \vec X)
=  \sum_{a=1}^A \pi_a \int d^N \vec X \ X_j  G^{[\vec \mu^{[a]}; {\bold \Sigma}^{[a]}]} (\vec X)
=   \sum_{a=1}^A \pi_a \mu^{[a]}_j
\label{mupia}
\end{eqnarray}
while the correlation matrix ${\bold C}^{ini} $ has for matrix elements
 \begin{eqnarray}
C_{i,j}^{ini} \equiv \int d^N \vec X \ X_i X_j P^{ini}( \vec X)
=  \sum_{a=1}^A \pi_a \int d^N \vec X \ X_i X_j  G^{[\vec \mu^{[a]}; {\bold \Sigma}^{[a]}]} (\vec X)
=   \sum_{a=1}^A \pi_a \left( \mu^{[a]}_i \mu^{[a]}_j +  \Sigma^{[a] }_{i,j} \right) 
\label{cijpia}
\end{eqnarray}
More generally, correlations of arbitrary order can be computed via the application of the Wick's theorem to each Gaussian distribution.


\subsubsection{ Properties of the forward Ornstein-Uhlenbeck propagator }

The forward propagator is the product of the $N$ elementary Ornstein-Uhlenbeck propagators of Eq. \ref{OUpropagator}
  \begin{eqnarray}
\langle \vec x \vert e^{  t W} \vert \vec X \rangle 
 = \prod_{n=1}^N \left( \sqrt {\frac{\omega}{\pi (1-e^{-2 \omega t} ) }} 
 e^{ \displaystyle - \omega \frac{ \left( x_n - X_n e^{-\omega t} \right)^2}{(1-e^{-2 \omega t} )} } \right)
 = \left(\frac{\omega}{\pi (1-e^{-2 \omega t} ) } \right)^{\frac{N}{2}}
 e^{ \displaystyle - \omega \frac{ \left( \vec x - \vec X e^{-\omega t} \right)^2}{(1-e^{-2 \omega t} )} } 
\label{OUpropagatorN}
\end{eqnarray}

The two first moments of the Gaussian steady state $p_*(x)$ of Eq. \ref{OUsteady}
 \begin{eqnarray}
m_1^* && =0
\nonumber \\
m_2^* && = \frac{1}{2 \omega}
\label{OUsteadymoments}
\end{eqnarray}
yield that the first eigenvector $l_1(x) $ of Eq. \ref{l1pearson} reduces to
\begin{eqnarray}
l_1(x) 
= \frac{ x - m_1^*}{\sqrt{ m_2^* - \left(m_1^*\right)^2 } } = x \sqrt{2 \omega} 
 \label{l1pearsonOU}
\end{eqnarray}
so that the convergence of Eq. \ref{factorizedpropagatorspectralonlyleftfirstexcitedN} 
  \begin{eqnarray}
 \langle \vec x \vert e^{  t W} \vert \vec X \rangle
&& = P_*( \vec x) \bigg[  1+  e^{- t e_1 } \sum_{n=1}^N l_1(x_n) l_1(X_n) +...\bigg]
\nonumber \\
&& =  \left( \frac{\omega}{\pi  } \right)^{\frac{N}{2}} e^{  - \omega \vec x^2 } 
\bigg[  1+  e^{- t \omega } 2 \omega \sum_{n=1}^N x_n X_n    +...\bigg]
=  \left( \frac{\omega}{\pi  } \right)^{\frac{N}{2}} e^{  - \omega \vec x^2 } 
\bigg[  1+  e^{- t \omega } 2 \omega \vec  x . \vec X    +...\bigg]
\label{factorizedpropagatorspectralonlyleftfirstexcitedNou}
\end{eqnarray}
is in agreement with the direct expansion of Eq. \ref{OUpropagatorN} as it should.


\subsubsection{ Properties of the forward solution $P(\vec x,t)$ }

The forward solution $P(\vec x,t)$ obtained from the application of the Gaussian propagator of Eq. \ref{OUpropagatorN}
to the Gaussian mixture $P^{ini}( \vec X)$ of Eq. \ref{mixture}
  \begin{eqnarray}
P(\vec x,t) = \int d^N \vec X \langle \vec x \vert e^{  t W} \vert \vec X \rangle P^{ini}(\vec X)
&&  =\sum_{a=1}^A \pi_a  \int d^N \vec X \left(\frac{\omega}{\pi (1-e^{-2 \omega t} ) } \right)^{\frac{N}{2}}
 e^{ \displaystyle - \omega \frac{ \left( \vec x - \vec X e^{-\omega t} \right)^2}{(1-e^{-2 \omega t} )} } 
 G^{[\vec \mu^{[a]}; {\bold \Sigma}^{[a]}]} (\vec X)
\nonumber \\
&&  \equiv \sum_{a=1}^A \pi_a \ \ G^{[\vec \mu^{[a]}(t); {\bold \Sigma}^{[a]}(t)]} (\vec x)
\label{mixturet}
\end{eqnarray}
is also a mixture of multivariate gaussian distributions
 \begin{eqnarray}
G^{[\vec \mu^{[a]}(t); {\bold \Sigma}^{[a]}(t)]} (\vec x) 
\equiv   \int d^N \vec X \left(\frac{\omega}{\pi (1-e^{-2 \omega t} ) } \right)^{\frac{N}{2}}
 e^{ \displaystyle - \omega \frac{ \left( \vec x - \vec X e^{-\omega t} \right)^2}{(1-e^{-2 \omega t} )} } 
 G^{[\vec \mu^{[a]}; {\bold \Sigma}^{[a]}]} (\vec X)
\label{mixtureta}
\end{eqnarray}
with the following time-dependent parameters $[\vec \mu^{[a]}(t); {\bold \Sigma}^{[a]}(t)] $.
The averaged values $ \mu^{[a]}_j(t) $ 
 \begin{eqnarray}
 \mu^{[a]}_j(t) && \equiv \int d^N \vec x x_j G^{[\vec \mu^{[a]}(t); {\bold \Sigma}^{[a]}(t)]} (\vec x) 
=   \int d^N \vec X   G^{[\vec \mu^{[a]}; {\bold \Sigma}^{[a]}]} (\vec X)
 \int d^N \vec x x_j \left(\frac{\omega}{\pi (1-e^{-2 \omega t} ) } \right)^{\frac{N}{2}}
 e^{ \displaystyle - \omega \frac{ \left( \vec x - \vec X e^{-\omega t} \right)^2}{(1-e^{-2 \omega t} )} }
 \nonumber \\
 && = \int d^N \vec X   G^{[\vec \mu^{[a]}; {\bold \Sigma}^{[a]}]} (\vec X)X_j e^{-\omega t} =  \mu^{[a]}_j e^{-\omega t} 
\label{mixturemuta}
\end{eqnarray}
decay exponentially as $e^{-\omega t} $ from their initial values $\mu^{[a]}_j $.
The correlations 
 \begin{eqnarray}
C_{i,j}^{[a]} (t) && \equiv \int d^N \vec x x_i x_j G^{[\vec \mu^{[a]}(t); {\bold \Sigma}^{[a]}(t)]} (\vec x) 
=   \int d^N \vec X   G^{[\vec \mu^{[a]}; {\bold \Sigma}^{[a]}]} (\vec X)
 \int d^N \vec x x_i x_j \left(\frac{\omega}{\pi (1-e^{-2 \omega t} ) } \right)^{\frac{N}{2}}
 e^{ \displaystyle - \omega \frac{ \left( \vec x - \vec X e^{-\omega t} \right)^2}{(1-e^{-2 \omega t} )} }
 \nonumber \\
 && =   \int d^N \vec X   G^{[\vec \mu^{[a]}; {\bold \Sigma}^{[a]}]} (\vec X)
 \left[ X_i e^{-\omega t} X_j e^{-\omega t} + \delta_{i,j} \frac{1- e^{-2 \omega t}}{2 \omega}  \right] 
  \nonumber \\
 && =   \left( \mu^{[a]}_i \mu^{[a]}_j +  \Sigma^{[a] }_{i,j} \right)  e^{- 2 \omega t} + \delta_{i,j} \frac{1- e^{-2 \omega t}}{2 \omega}  
\label{cijat}
\end{eqnarray}
lead to the covariance matrix elements
 \begin{eqnarray}
\Sigma_{i,j}^{[a]} (t) = C_{i,j}^{[a]} (t) - \mu^{[a]}_i(t) \mu^{[a]}_j(t) 
  =     \Sigma^{[a] }_{i,j}   e^{- 2 \omega t} + \delta_{i,j} \frac{1- e^{-2 \omega t}}{2 \omega}  
\label{sigmaijat}
\end{eqnarray}
where the first term represents the exponential decay as $e^{- 2 \omega t} $ from the initial values $\Sigma^{[a] }_{i,j}  $,
while the second term representing the effect of the noising dynamics
is diagonal and grows from zero towards the asymptotic value $\frac{1}{2 \omega} $.

The asymptotic behavior of Eq. \ref{factorizedpropagatorspectralonlyleftfirstexcitedNou} for the propagator
leads to the following asymptotic behavior for the forward solution
  \begin{eqnarray}
P(\vec x,t) = \int d^N \vec X \langle \vec x \vert e^{  t W} \vert \vec X \rangle P^{ini}(\vec X)
&&  =  \left( \frac{\omega}{\pi  } \right)^{\frac{N}{2}} e^{  - \omega \vec x^2 } 
\bigg[  1+  e^{- t \omega } 2 \omega \sum_{n=1}^N x_n \int d^N \vec X X_n  P^{ini}(\vec X)    +...\bigg]
\nonumber \\
&&  =  \left( \frac{\omega}{\pi  } \right)^{\frac{N}{2}} e^{  - \omega \vec x^2 } 
\bigg[  1+  e^{- t \omega } 2 \omega \sum_{n=1}^N x_n \mu_n^{ini}    +...\bigg]
\label{forwardsolutionfirstexcitedNou}
\end{eqnarray}
that involves the initial averaged values $\mu_n^{ini} $ of Eq. \ref{mupia}.


\subsubsection{ Convergence properties of the reconstructive backward dynamics  }

For any finite time $t$,
the backward propagator of Eq. \ref{backwardpropagator}
has in general no simple properties because the Gaussian mixture $P(\vec x,t) $ of Eq. \ref{mixturet} appears in the denominator
  \begin{eqnarray}
 B(  \vec X, 0 \vert \vec x, t ) && = \frac{ \langle \vec x \ \vert e^{  t W} \vert \vec X \rangle P^{ini}(\vec X) }{  P( \vec x, t) }
 \nonumber \\
 && =  \frac{\displaystyle \sum_{a=1}^A \pi_a   \left(\frac{\omega}{\pi (1-e^{-2 \omega t} ) } \right)^{\frac{N}{2}}
 e^{  - \omega \frac{ \left( \vec x - \vec X e^{-\omega t} \right)^2}{(1-e^{-2 \omega t} )} } 
 G^{[\vec \mu^{[a]}; {\bold \Sigma}^{[a]}]} (\vec X) }
 { \displaystyle \sum_{a'=1}^A \pi_{a'} \ \ G^{[\vec \mu^{[a']}(t); {\bold \Sigma}^{[a']}(t)]} (\vec x) }
\label{backwardpropagatormixture}
\end{eqnarray}
so that we will now focus only the asymptotic properties for large time $t \to + \infty$.
The convergence of Eq. \ref{B0reconstructedbackwardspectraldifferenceleadingOUfinal}
for the reconstructive backward dynamics 
involves the averaged values $\mu_n^{ini}$ of Eq. \ref{mupia}
 \begin{eqnarray}
B^{[0,t]}(\vec X) - P^{ini}(\vec X)
&&   \opsimeq_{t \to + \infty}   
   e^{- t 2 e_1 } 
    P^{ini}(\vec X)   
\sum_{n=1}^N   \frac{\left[  \int d^N \vec Y Y_n P^{ini} (\vec Y)- m_1^* \right]
 \left[   \int d^N \vec Z Z_n P^{ini} (\vec Z)- X_n \right] }{m_2^* - \left(m_1^*\right)^2  }   
    +...
    \nonumber \\
    &&  \opsimeq_{t \to + \infty}  2 \omega   e^{- t 2 \omega } 
    P^{ini}(\vec X)   \sum_{n=1}^N   \mu_n^{ini} \left[   \mu_n^{ini} - X_n \right]    
\label{B0reconstructedbackwardspectraldifferenceleadingOUfinal1d}
\end{eqnarray}

In particular, the convergence of the averaged value of $X_j$ concerning the single pixel $j$
  \begin{eqnarray}
 \int d^N \vec X \ X_j \left[ B^{[0,t]}(\vec X) -  P^{ini}(\vec X) \right]
    &&  \opsimeq_{t \to + \infty}  2 \omega   e^{- t 2 \omega } 
       \sum_{n=1}^N   \mu_n^{ini} \left[   \mu_n^{ini} \int d^N \vec X \ X_j P^{ini}(\vec X) - \int d^N \vec X \ X_j X_n P^{ini}(\vec X)\right]
       \nonumber \\
  &&  \opsimeq_{t \to + \infty}  2 \omega   e^{- t 2 \omega } 
       \sum_{n=1}^N   \mu_n^{ini} \left[   \mu_n^{ini} \mu_j^{ini}  - C^{ini}_{jn} \right]          
\label{B0reconstructedbackwardspectraldifferenceleadingOUfinal1dXj}
\end{eqnarray}
involves the averaged values $\mu_n^{ini} $ of Eq. \ref{mupia}
and the correlations $C^{ini}_{jn} $ of Eq. \ref{cijpia}.


\subsubsection{ Special case $\pi_a=\frac{1}{A}$ and ${\bold \Sigma}^{[a]}=0 $ when $P^{ini}(\vec X)$ is the empirical distribution of the $A$ images $\vec \mu^{[a]} $ }

For the special case of equiprobable weights $\pi_a=\frac{1}{A}$ and vanishing covariances matrices ${\bold \Sigma}^{[a]}=0 $, the mixture of multivariate Gaussian distributions of Eq. \ref{mixture}
reduces to the empirical distribution of the $A$ images $\vec \mu^{[a]} $
 \begin{eqnarray}
P^{ini}_{empi}( \vec X) =\frac{1}{A}  \sum_{a=1}^A  G^{[\vec \mu^{[a]}; {\bold \Sigma}^{[a]}=0]} (\vec X)
= \frac{1}{A} \sum_{a=1}^A \delta^{(N)} \left( \vec X - \vec \mu^{[a]} \right)
\label{empi}
\end{eqnarray}
that represents the continuous analog of Eq. \ref{empiC}.

Then the asymptotic behavior of Eq. \ref{B0reconstructedbackwardspectraldifferenceleadingOUfinal1d}
 \begin{eqnarray}
B^{[0,t]}(\vec X) 
    &&  \opsimeq_{t \to + \infty}  P^{ini}_{empi}(\vec X) \left( 1 + 2 \omega   e^{- t 2 \omega } 
      \sum_{n=1}^N   \mu_n^{ini} \left[   \mu_n^{ini} - X_n \right]   \right) 
      \nonumber \\
      &&  \opsimeq_{t \to + \infty}   
      \sum_{a=1}^A \delta^{(N)} \left( \vec X - \vec \mu^{[a]} \right) 
       \frac{1}{A} \left( 1 + 2 \omega   e^{- t 2 \omega } 
      \sum_{n=1}^N  \left[ \frac{1}{A}  \sum_{a'=1}^A  \mu^{[a']}_n\right] \left[ \frac{1}{A}  \sum_{a''=1}^A  \mu^{[a'']}_n - \mu^{[a]}_n \right]   \right) 
\label{B0reconstructedbackwardspectraldifferenceleadingOUfinal1dempi}
\end{eqnarray}
concerns the convergence of the weights of the $A$ images $\vec \mu^{[a]} $ towards their equiprobable values $\frac{1}{A}$, in agreement with the general discussion of Eqs \ref{B0reconstructedbackwarddifferenceleadingempi}
and \ref{backwardpropagatorempiwa}.


\section{ Conclusion }

\label{sec_conclusion}

In summary, we have analyzed the convergence properties of Markov models for image generation
as a function of the time-window $[0,t]$ used to implement the forward noising dynamics and the backward denoising dynamics.

We have first described how the eigenvalues $e_{\alpha}$ and the left eigenvectors $l_{\alpha}$
of the forward Markov generator $w$ governing the dynamics of a single pixel
are useful to understand how the non-trivial properties of the initial condition $P^{ini}(\vec C)$
involving the $N$ pixels
are gradually destroyed during the time-window $[0,t]$.

We have then explained how the spectral properties of the forward dynamics
are also useful to characterize the convergence properties of the reconstructive backward dynamics.

Finally, we have applied this general framework to two specific cases : 

(a) when each pixel has only two states $S_n=\pm 1$ with Markov jumps between them, 
we have explained how the global left eigenvectors have very natural interpretations
 in terms of the magnetizations and of the multi-spin-correlations;
 as illustrative example, we have described the case where the initial distribution $P^{ini}(S_1,S_2,..,S_N) $
of the $N$ spins corresponds to the Boltzmann distribution 
of the one-dimensional Ising model involving arbitrary couplings $J_{n=1,..,N}$;

(b) when each pixel is characterized by a continuous variable $x_n$ that diffuses on an interval $]x_-,x_+[$,
we have explained how the left eigenvectors have very natural interpretations in terms of polynomials of $x_n$
whenever the diffusion process belongs to the Pearson 'half-family' with finite steady moments, 
and we have given the three following examples:

(i) the Jacobi process on the finite interval $x \in ]0,1[$;

 (ii) the Square-Root or Cox-Ingersoll-Ross process on the positive axis $x \in ]0,+\infty[$ ;

 (iii) the Ornstein-Uhlenbeck process on the whole axis $x \in ]-\infty,+\infty[$, which is
 the most standard synamics used in the field of Generative diffusions models;
 as illustrative exanple, we have described the case where the initial distribution $P^{ini}(\vec X)$ is a mixture of multivariate Gaussian distributions or of delta functions.


\end{document}